\documentclass[a4paper,11pt]{article}
\usepackage{jheppub}
\usepackage[T1]{fontenc}
\usepackage{amsmath,xcolor}
\usepackage{amsmath, amssymb,ascmac,fancybox,mathrsfs,bbold}
\usepackage{color,booktabs,fancyhdr,bm,comment,cancel}
\usepackage{hyperref,appendix}
\usepackage[]{graphicx,graphics}
\usepackage{mathtools}
\usepackage{appendix}
\usepackage{tikz}

\usetikzlibrary{positioning}
\usetikzlibrary{arrows}
\usetikzlibrary{positioning,decorations.pathmorphing,decorations.markings,arrows}

\usetikzlibrary{math}

\makeatletter
\renewcommand{\p@subsection}{}
\makeatother

\makeatletter
\newcommand{\xequal}[2][]{\ext@arrow 0055{\equalfill@}{#1}{#2}}
\def\equalfill@{\arrowfill@\Relbar\Relbar\Relbar}
\makeatother

\renewcommand{\thesection}{\arabic{section}}
\renewcommand{\thesubsection}{\arabic{section}.\arabic{subsection}}

\makeatletter

\@addtoreset{equation}{section} 
\makeatother

\newcommand{\chushi}[1]{ }

\newcommand{\C}{\mathcal{C}}
\newcommand{\uu}{\mathfrak{u}}
\newcommand{\ww}{\mathfrak{w}}

\newcommand{\Gr}{\mathrm{Gr}}
\newcommand{\SpGr}{\mathrm{SpGr}}
\newcommand{\GL}{\mathrm{GL}}
\newcommand{\SL}{\mathrm{SL}}
\newcommand{\SU}{\mathrm{SU}}
\newcommand{\U}{\mathrm{U}}
\newcommand{\SO}{\mathrm{SO}}
\renewcommand{\tilde}{\widetilde}

\numberwithin{equation}{section}
\let\calccommentout\iffalse 
\let\calcshow\iftrue 

\newcommand{\p}{\partial}

\newcommand {\mathsym}[1]{{}}
\newcommand {\unicode}[1]{{}}

\allowdisplaybreaks

\newcommand{\dd}{\mathrm{d}}
\newcommand{\adot}{\dot{a}}
\newcommand{\bdot}{\dot{b}}
\newcommand{\aldot}{\dot{\alpha}}
\newcommand{\bedot}{\dot{\beta}}
\newcommand{\psix}{\overset{\scriptscriptstyle{(6)}}{p}{}}
\newcommand{\lambsix}{\overset{\scriptscriptstyle{(6)}}{\lambda}{}}
\newcommand{\lambtsix}{\overset{\scriptscriptstyle{(6)}}{\widetilde{\lambda}}{}}
\newcommand{\lambar}{\overline{\lambda}}
\newcommand{\lambtbar}{\widetilde{\overline{\lambda}}}
\newcommand{\barp}{\overline{p}}

\newcommand{\bari}{\overline{i}{}}

\newcommand{\tildu}{\tilde{u}}

\newcommand{\si}{\mathsf{I}}
\newcommand{\sj}{\mathsf{J}}
\newcommand{\sa}{\mathsf{A}}
\newcommand{\sbb}{\mathsf{B}}
\newcommand{\etabar}{\widetilde{\eta}{}}
\newcommand{\parbar}{\widetilde{\partial}{}}
\newcommand{\qsix}{\overset{\scriptscriptstyle{(6)}}{Q}{}}
\newcommand{\qtildesix}{\overset{\scriptscriptstyle{(6)}}{\widetilde{Q}}{}}

\newcommand{\ang}[1]{\langle #1 \rangle}
\newcommand{\bd}{\dot{b}}
\newcommand{\ad}{\dot{a}}
\newcommand{\tc}{\tilde{c}}

\author[a]{Veronica Calvo Cortes,}
\emailAdd{veronica.calvo@mis.mpg.de}
\affiliation[a]{Max Planck Institute for Mathematics in the Sciences, Leipzig, Germany.}
\author[b]{Yassine El Maazouz,}
\emailAdd{maazouz@caltech.edu}
\affiliation[b]{California Institute of Technology, Pasadena, California.}
\author[c]{Subramanya Hegde,}
\emailAdd{subbu@mpp.mpg.de}
\affiliation[c]{Max Planck Institute for Physics, Munich, Germany}
\author[d,e]{Amit Suthar}
\emailAdd{amitsuthar@imsc.res.in}
\affiliation[d]{The Institute of Mathematical Sciences, CIT Campus, Taramani, Chennai 600113, India.}
\affiliation[e]{Homi Bhabha National Institute, Training School Complex, Anushakti Nagar, Mumbai 400094, India}

\title{Symplectic Grassmannian description of the Coulomb branch three and four point amplitudes}

\begin{document}
	\count\footins = 1000 

        \abstract{We present a formulation of the three- and four-point amplitudes on the Coulomb branch of $\mathcal{N}=4$ SYM as integrals over the symplectic Grassmannian. We demonstrate that their kinematic spaces are equivalent to symplectic Grassmannians $\SpGr(n,2n)$. For the three-point case, we express the amplitude as an integral over the symplectic Grassmannian in a specific little group frame. In the four-point case, we show that the integral yields the amplitude up to a known kinematic factor. Building on the four-dimensional analysis, we also express the six-dimensional $\mathcal{N} = (1,1)$ SYM amplitude in terms of four-dimensional variables in a form that makes its symplectic Grassmannian structure manifest.}

	\maketitle
	
	\newpage

\section{Introduction}

Over the past two decades, significant advances have been made in the efficient computation of scattering amplitudes, bypassing the complexity of traditional Feynman diagrammatics \cite{Dixon:1996wi,Elvang:2013cua,Henn:2014yza}. Central to this progress are on-shell methods that exploit the factorization properties of amplitudes and loop integrands when internal propagators are placed on-shell. A foundational breakthrough in this direction came from Britto, Cachazo, Feng, and Witten (BCFW) \cite{Britto:2004ap, Britto:2005fq} who introduced a recursion relation for tree-level gluon amplitudes thereby revealing the underlying simplicity first observed by Parke and Taylor \cite{Parke:1986gb}. The BCFW method was also instrumental in establishing recursion relations for tree-level amplitudes in a variety of other quantum field theories \cite{Luo:2005rx,Cheung:2008dn,Elvang:2008na,Feng:2009ei,Cheung:2010vn}. These tree-level techniques were subsequently extended to loop integrands through the development of unitarity-based methods \cite{Bern:1994cg,Bern:1994zx,Bern:1995db,Bern:1996ja,Bern:1996je,Bern:1996fj,Bern:1997nh,Bern:1997sc,Bern:1998ug,Bern:2004cz,Britto:2004nc,Bern:2011qt}.

The on-shell methods discussed above are primarily suited for massless particles and rely on massless spinor-helicity variables \cite{Berends:1981rb}. In this context, planar maximally supersymmetric $(\mathcal{N}=4)$ Yang-Mills theory in four dimensions serves as an ideal setting for applying such techniques. At tree level, all possible BCFW shifts are valid in this theory, and at loop level, it admits a simplified basis of master integrals — attributes that have earned it the reputation of being "the simplest quantum field theory" \cite{Arkani-Hamed:2008owk}. The striking properties of the amplitudes in this theory led to the discovery of a hidden dual conformal symmetry \cite{Drummond:2008vq}, which was later understood to be part of a larger Yangian symmetry structure \cite{Drummond:2009fd}.

The search for a mathematical framework that makes the Yangian invariance of $ \mathcal{N}=4 $ SYM amplitudes manifest led to a geometric reformulation. In this approach, perturbative amplitudes in $ \mathcal{N}=4 $ SYM were expressed as integrals over Grassmannians $\mathrm{Gr}(k,n)$, which parametrize $k$-planes in an $ n $-dimensional space~\cite{Arkani-Hamed:2009ljj,Arkani-Hamed:2009kmp,Arkani-Hamed:2009nll,Arkani-Hamed:2009pfk,Arkani-Hamed:2012zlh}. The use of Grassmannians renders (super)momentum conservation linear in the external kinematic variables, greatly simplifying the construction of larger ``on-shell functions'' by gluing together three-particle amplitudes, with all internal legs placed on-shell. Concurrently, it was discovered that the Next-to-Maximal Helicity Violating (NMHV) amplitudes in pure Yang-Mills theory are proportional to volumes of certain polytopes in $\mathbb{CP}^3$~\cite{Hodges:2009hk}. These insights, combined with the Grassmannian formulation, culminated in the discovery of the \emph{Amplituhedron}~\cite{Arkani-Hamed:2013jha}---a geometric object that encapsulates scattering amplitudes and suggests that fundamental principles like locality and unitarity emerge from the deeper geometry of total positivity.

While the rich structure of $ \mathcal{N}=4 $ SYM has provided deep insights into scattering amplitudes in massless theories~\cite{Henn:2020omi}, the study of massive amplitudes remains comparatively less explored. Early efforts to dealing with massive particles involved expressing massive momenta as a sum of two massless momenta, allowing the use of massless spinor-helicity variables~\cite{Kleiss:1985yh,Kleiss:1988xr}. More recently, a formalism based on little-group covariant massive spinor-helicity variables was introduced in~\cite{Arkani-Hamed:2017jhn}, offering a more natural treatment of massive states. A particularly tractable example featuring massive amplitudes is provided by the Coulomb branch of $ \mathcal{N}=4 $ SYM, which we describe below.

The $\mathcal{N}=4$ SYM theory has six scalars, and they can acquire a vacuum expectation value, resulting in the Coulomb branch. In this theory, the gauge symmetry breaks into $\U(N) \to \prod_i \U(N_i)$, where $\sum_i N_i = N$ is the rank of the gauge group\footnote{Note that different number of scalars can be given vacuum expectation values to obtain the same gauge group breaking, resulting in the Coulomb branch theory preserving different amount of $R$--symmetry. We will consider the set-up where a single scalar acquires a vacuum expectation value which leaves an unbroken ${\rm USp}(4)$ $R$--symmetry along the lines of \cite{Alday:2009zm,Kiermaier:2011cr}. More recently, authors of \cite{Arkani-Hamed:2023epq,Flieger:2025ekn} have studied a different set up where two independent scalars are given a vacuum expectation value, in the context of deforming the amplituhedron geometry to obtain massive amplitudes.}. 
The on-shell degrees of freedom stay the same, but the components rearrange into massive spin-1 $W$-bosons and their supersymmetric partners. 
To name a few, the amplitudes for this theory were initially studied in \cite{Alday:2009zm,Boels:2010mj,Craig:2011ws, Kiermaier:2011cr, Plefka:2014fta,Caron-Huot:2014gia} where various nice properties of these amplitudes were found, such as dual conformal invariance \cite{Alday:2009zm}, and the absence of triangle master integrals at loop level \cite{Boels:2010mj}. 
The massive spinor-helicity variables due to \cite{Arkani-Hamed:2017jhn} helped write the amplitudes in little group covariant form \cite{Herderschee:2019dmc, Herderschee:2019ofc}. 
In \cite{Herderschee:2019dmc}, the tree-level Coulomb branch amplitudes were written, where the authors found that these amplitudes are super-BCFW constructible. 
In \cite{Abhishek:2023lva}, some loop-level amplitudes were calculated using the unitarity cut methods. 
In \cite{MdAbhishek:2023nvg}, the same loop amplitudes were calculated using generalized unitarity, which led naturally to the on-shell functions for the Coulomb branch of $\mathcal{N}=4$. 
In particular, the authors performed the BCFW bridge construction in spinor-helicity variables, which provides a natural connection between the on-shell functions and the amplitudes. 
The various remarkable properties described above call for an investigation for a Grassmannian like formulation of these on-shell functions. 
In this paper we carry out such an investigation, focusing on three and four point amplitudes on the Coulomb branch.

In the massless case, the amplitudes of the $\mathcal{N}=4$ SYM are categorized in different helicity violating sectors $N^{k-2}MHV$, $k=2$ being the MHV case. Amplitudes in $N^{k-2}MHV$ sector are written as integrals over $\Gr(k,n)$ \cite{Arkani-Hamed:2012zlh}:
\begin{align*}
     \mathcal{I}_{n,k} &= \int \frac{d^{n\times k}C}{\GL(k)}\frac{1}{\prod M_i} \delta^{2k}\big(C.\Lambda^T\big)\,\delta^{2(n-k)}\,\big(C^{\perp}.\tilde{\Lambda}^T\big)\,\delta^{4k}\big(C.\eta^T\big) ~,
\end{align*}
and are related to a particular Amplituhedron $\mathcal{A}_{k,n}^{(L)}$. For the massive particles, a fixed helicity is no longer a Lorentz invariant notion and different $k$ sectors end up mixing amongst each other. This can be called the \emph{unification in the IR} \cite{Arkani-Hamed:2017jhn}. Upon taking the high energy limit, the masses can be ignored and we end up having $N^{k-2}MHV$ amplitudes with their $Gr(k,n)$ description. Based on earlier works \cite{Cachazo:2018hqa,Schwarz:2019aat,Bering:2022tdr}, it is expected that the Coulomb branch amplitudes should be written as integrals over a symplectic Grassmannian $\SpGr(n,2n)$.  
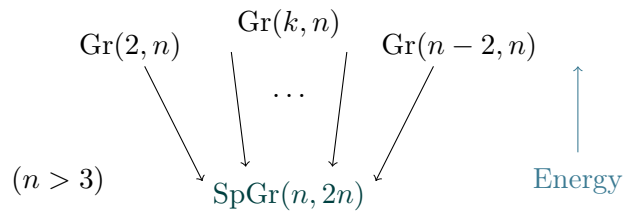
\begin{center}
        \begin{tikzpicture}[scale=0.38]
        \draw[->] (-5,4) -- (-3,0);
        \draw[->] (5,4) -- (3,0);
        \draw[->] (-2,4.5) -- (-1.5,0.5);
        \draw[->] (2,4.5) -- (1.5,0.5);
        \node[teal!50!black] at (0,-0.5) {$\SpGr(n,2n)$};
        \node at (0,3) {$\hdots$};
        \node at (-5.5,4.7) {$\Gr(2,n)$};
        \node at (5.9,4.7) {$\Gr(n-2,n)$};
        \node at (0,5.5) {$\Gr(k,n)$};
        \draw[->, cyan!50!black] (10,1) -- (10,4);
        \node[cyan!50!black] at (10,0) {Energy};
        \node at (-8,0) {$(n>3)$};
    \end{tikzpicture}
    \end{center}
For three particles, we have $\SpGr(3,6) \to \Gr(1,3) \ \text{or} \ \Gr(2,3)~.$ For the massless on-shell functions of $\mathcal{N}=4$, the basic building blocks are the MHV ($\Gr(2,3)$) and anti-MHV ($\Gr(1,3)$) amplitudes, which are referred to as the black and white dots. The massive analog can be called the `gray' dot, which becomes black or white in the massless limit. In this paper, we have written a $\SpGr(3,6)$ integral for such a gray dot. This Grassmannian description leads to a geometric understanding of the special three-body kinematics. Further, we have also written an integral over a symplectic Grassmannian $\SpGr(4,8)$ that leads to the four point Coulomb branch amplitude up to a kinematic factor. 

The massive supermultiplet of $\mathcal{N}=4$ in four dimensions is the same as the massless supermultiplet of $\mathcal{N}=(1,1)$ in six dimensions. We can interpret the mass of the four-dimensional particles as the momentum in the compact dimensions. In appendix \ref{Appendix:B dim reduction of spinor-helicity variables}, we work out the embedding of massive spinor-helicity variables of four dimensions \cite{Arkani-Hamed:2017jhn} in the massless spinor-helicity variables of six dimensions \cite{Cheung:2009dc}, in such a way that there is no loss of kinematic data during dimensional reduction. The similarity in the two theories makes it possible to see the underlying symplectic Grassmannian for the six-dimensional theory, and it matches well with past work \cite{Bering:2022tdr}. 

\subsection{Main results}
The three-point Coulomb branch super-amplitude can be written merely as a supercharge-conserving delta function as follows:
\begin{align} \label{three-point amplitude written in C_*}
            \mathcal{A}_3 = \delta^6\left(C_*.\Omega.\eta^T\right) ~~, \hspace{1cm} C_* = \begin{tikzpicture}[scale=0.5, baseline={([yshift=-.5ex]current bounding box.center)}]
        \begin{scope}[transparency group]
        \begin{scope}[blend mode=multiply]
            \fill[fill=red!40, rounded corners, opacity=0.5] (-3.1,-0.52) rectangle (3.3,1.52);
            \fill[fill=cyan!50, rounded corners, opacity=0.5] (-3.3,-1.52) rectangle (3.1,0.52);
            \node at (0,0) {$ \begin{matrix}
                \langle q1^I\rangle & \langle q2^I\rangle & \langle q3^I\rangle \\
                \langle u1^I\rangle & \langle u2^I\rangle & \langle u3^I\rangle \\
                [q1^I] & [q2^I] & [q3^I]
            \end{matrix} $};
        \end{scope}
        \end{scope}
        \node[red!80!black] at (4,0.5) {$\bigg\}\tilde{\Lambda} $};
        \node[cyan!50!black] at (-4,-0.5) {$\Lambda \bigg\{$}; 
    \end{tikzpicture} ~~. 
\end{align} 
The special three-body kinematics originates from the spaces $\Lambda$ and $\tilde{\Lambda}$ having an intersection, parameterized by the $u$ variables: $\langle ui^I\rangle = - [ui^I]~.$ The above form of the amplitude makes the $\SL(3)$ invariance $(C_* \to M.C_*)$ manifest, and it satisfies $C_*.\Omega.C_*^T = 0$, hinting at the underlying $\SpGr(3,6)$.

The four-particle amplitude can be written as follows:
\begin{align*}
    \mathcal{A}_4 = \frac{1}{(1^12^11^22^2)(2^13^12^23^2)}\delta^8\left(C_*.\Omega.\eta^T\right)~~, \qquad \quad C_* = \left( \begin{matrix} \Lambda_{2\times 8} \\ \tilde{\Lambda}_{2\times 8} \end{matrix} \right)~.
\end{align*}
The $\SL(4)$ invariance and $C_*.\Omega.C_*^T = 0$ hints at $\SpGr(4,8)~.$ The coefficient in front of the delta function is written in terms of minors of $C_*$.

The $n$ point on-shell function or amplitude for the Coulomb branch is written as an integral over the symplectic Grassmannian as follows:
\begin{equation} \label{actual integral in intro}
    {F}_n = \int \frac{\dd^{n\times 2n}C}{\GL(n)}\delta^{n(n-1)/2}\big(C.\Omega.C^T\big)\ f_n\left(\{M_i\}\right)\ \delta^{2n}\left(C.\Omega.\Lambda^T\right)\, \delta^{2n}\left(C.\Omega.\tilde{\Lambda}^T\right)\, \delta^{2n}\left(C.\Omega.\eta^T\right) ~.
\end{equation}
The functions $f_n$ are functions of the minors of $C$. For general $n$, they are not known. For $n=3$, multiple choices of $f_3$ give the correct three-particle amplitude. For instance, one of them is given by:
\begin{align*}
    f_3 = \frac{1}{(1^u2^w3^w)(2^u3^w1^w)} + \frac{1}{(2^u3^w1^w)(3^u1^w2^w)} + \frac{1}{(3^u1^w2^w)(1^u2^w3^w)}~.
\end{align*}
The three brackets $(i^uj^wk^w)$ are minors of the $C$ matrix, written in a specific (massive) little group frame dictated by the special three-body kinematics. For $n=4$, the following (non-unique) choice of $f_4$ gives the correct four-point amplitude:
\begin{align*}
    f_4 = \frac{s_{12}s_{23}}{-2s_{13}}\left(\frac{1}{(1^11^22^12^2)(2^12^23^13^2)(3^13^24^14^2)} + \text{cyclic permutations}\right)~.
\end{align*}
Note it has explicit external kinematic dependence.

The integral over a symplectic Grassmannian is better suited for $\mathcal{N}=(1,1)$ SYM in six dimensions. We find how the massless $d=6$ spinor-helicity variables decompose into the massive $d=4$ spinor-helicity variables, and study the six-dimensional theory in variables suited for four dimensions. We express the $\mathcal{N}=(1,1)$ amplitudes as follows hinting at the underlying $\SpGr(n,2n)$. 
\begin{align} \label{three points six dimensional}
    \mathcal{A}_3 &= \delta^3\left(C_*.\Omega.\eta^T\right)\ \delta^3\left(\overline{C}_*.\Omega.\tilde{\eta}^T\right)~. \\
    \mathcal{A}_4 &= \frac{1}{(1^11^22^12^2)(2^12^23^13^2)}\delta^4\left(C_*.\Omega.\eta^T\right)\,\delta^4\left(\overline{C}_*.\Omega.\etabar^T\right)~~. \label{four point six dimensional}    
\end{align}

\subsection*{Outline of the paper}
Our article is structured as follows. 
In Section \ref{sec:2 review}, we briefly review the Coulomb branch amplitudes, focusing on the three-particle amplitude and the special three-body kinematics. Appendix \ref{Appendix-Review-Onshell} contains a review of half-BPS states on the Coulomb branch of $\mathcal{N}=4$ SYM. 
In Section \ref{sec:3 three-point amplitude}, we analyze the three-point amplitude and rewrite the supersymmetric delta functions in a suggestive form \eqref{three-point amplitude written in C_*}, hinting at the underlying $\SpGr(3,6)$. 
We discuss that momentum and mass conservation enforce a one-dimensional intersection space between $\Lambda$ and $\tilde{\Lambda}$, resulting in the three-body special kinematics. 
We also discuss the massless limit of the amplitude. In Section \ref{sec:4 symplectic grassmannian integrals}, we write an ansatz for the symplectic Grassmannian integral \eqref{actual integral in intro} and find the appropriate integrands for three and four-point amplitudes. Some of the detailed calculations are in Appendix \ref{Appendix:D evaluation of the grassmannian integrals}.
In Section \ref{sec:5 six dimensional theory}, we discuss the amplitudes of six-dimensional maximal SYM, and write them in a suggestive form \eqref{three points six dimensional} and \eqref{four point six dimensional}, hinting at $\SpGr(n,2n)$. Appendices \ref{Appendix:B dim reduction of spinor-helicity variables} and \ref{Appendix:C dim reduction of susy algebra} contain the appropriate decomposition of six-dimensional spinor-helicity variables and Grassmann $\eta$ variables into the four-dimensional ones. In Appendix \ref{Appendix:E equivalent of three-point amplitude}, we show in detail the equivalence between the new representation of the three-point amplitude presented here \eqref{three points six dimensional} and the presentation in \cite{Dennen:2009vk}.
In Section \ref{sec:6 discussion}, we conclude the article by summarizing our results, and discussing future directions.

\section{Review of the Coulomb branch three-point amplitude}
\label{sec:2 review}

In this section, we will review the three-point amplitude in the Coulomb branch of $\mathcal{N}=4$ SYM. 
Even though a generic all massive three-point amplitude does not have any special kinematics \cite{Arkani-Hamed:2017jhn}, the central charge conservation of the Coulomb branch introduces an additional condition on the kinematics as reviewed in Appendix-\ref{Appendix-Review-Onshell}. 
This selects a preferred little group frame, and the amplitudes are most naturally described in this little group frame. 
One of the main results of this paper is elucidating the geometry of this three-particle special-BPS kinematics in terms of a symplectic Grassmannian. 
In preparation for this, we will now recall the construction of the three-point amplitude as developed in~\cite{Herderschee:2019dmc}. 

Consider a three-particle kinematics where all the particles are massive and their masses satisfy the following condition\footnote{We have analytically continued the masses such that some masses are negative. The precise way in which this is implemented for the later sections is outlined in the appendix.}:
\begin{align}\label{3pt-mass-cons}
m_1+m_2+m_3=0.
\end{align}
When combined with momentum conservation and on-shell conditions, one obtains, 
\begin{align*}
2p_i\cdot p_j + 2m_im_j=\det([i^Ij^J]+\langle i^Ij^J\rangle)=0, \, \qquad \text{for any } 1 \leq i,j \leq 3.
\end{align*} 
The condition above implies that the $2 \times 2$ matrices inside the determinant have at most rank one. Therefore, one can write
\begin{equation}
\begin{aligned}
[1^I2^J] + \langle 1^I 2^J\rangle &=u_1^Iv_2^J\\
[2^J3^K] + \langle 2^J 3^K\rangle &=u_2^Jv_3^K\\
[3^K1^I] + \langle 3^K 1^I\rangle &=u_3^Kv_1^I.
\end{aligned}    
\end{equation}
Note that these equations are not independent. For instance,
\begin{equation*}
\begin{aligned}
u_1^I\left(v_2^Ju_{2J}\right)v_3^K&=([1^I2^J]+ \langle 1^I2^J\rangle)([2_J3^K]+ \langle 2_J 3^K\rangle)\\
&=\left(-\sum_i m_i\right) [1^I3^K]+ \left(\sum_i m_i\right) \langle 1^I3^K\rangle\\
&=0,
\end{aligned}
\end{equation*}
where from the first to the second line we have used spin sums and momentum conservation.
We can then argue: $u_i^I\propto v_i^I$. Further, from gauge fixing the tiny groups of each pair of particles we can set $u_i^I=v_i^I$. An important consequence is that we can define two special spinors in terms of spinor-helicity variables of any of the external particles and the $u_i^I$:
\begin{align*}
|u] \equiv u_{iI}|i^I],\nonumber \quad \text{and} \quad |u\rangle \equiv u_{iI}|i^I\rangle.
\end{align*}
The two spinors above are related by
\begin{align*}
\frac{p_i}{m_i}|u\rangle=-|u],\, \forall i.
\end{align*}
We can write the inverse relations as
\begin{align*}
u_i^I=-\frac{[ui^I]}{m_i}=\frac{\langle ui^I\rangle}{m_i}.
\end{align*}
The existence of these variables for three points puts constraints on the three-point super-amplitude. As we review in the appendix, for $n\ge 4$ point tree-level super-amplitudes, we have the super-momentum conserving delta functions $\delta^{(4)}(Q_{a+2})$ and $\delta^{(4)}(Q^{\dagger a})$ that appear as multiplicative factors. We can decompose these delta functions along two independent $\SL(2,\mathbb{C})$ directions:
\begin{equation}
\begin{aligned}
\delta^{(4)}(Q_{a+2})&=\frac{1}{[qr]^2}\delta^{(2)}([qQ])\delta^{(2)}([rQ]),\\
\delta^{(4)}(Q^{\dagger a})&=\frac{1}{\langle qr\rangle ^2}\delta^{(2)}(\langle qQ^\dagger\rangle)\delta^{(2)}(\langle rQ^\dagger\rangle),
\end{aligned}    
\end{equation}
where $[qr], \langle qr \rangle \neq 0$. For three points we can project, in particular, along $|u], |u\rangle$ directions and $|q], |q\rangle$ reference spinor directions such that $[qu], \langle qu \rangle \neq 0$. Then, due to the BPS condition satisfied by the super-charges and the three-point special kinematics discussed above, one obtains the following relation 
\begin{equation}
    \begin{aligned}
\langle u Q^{\dagger a} \rangle = - [uQ_{a+2}].
\end{aligned}
\end{equation}
Therefore, if we impose supersymmetry conservation along $\langle u Q^{\dagger a} \rangle$ then the supersymmetry along $[uQ_{a+2}]$ is automatically established. Hence, for the three-point super amplitude one has six super-momentum conserving delta functions instead of eight. One then needs to determine the overall kinematic factor, which was determined by using the massless limit in \cite{Herderschee:2019dmc}. The three-point super-amplitude is then written as,
\begin{equation}\label{3-point-amp}
\begin{aligned}
A^{(3)}&=\frac{1}{\langle q| p_1p_3|q\rangle}\delta^{(4)}(Q)\delta^{(2)}(\langle qQ^\dagger\rangle)\\
&=\frac{1}{[q|p_1p_3|q]}\delta^{(4)}(Q^{\dagger})\delta^{(2)}([qQ]).
\end{aligned}    
\end{equation}
The three-particle special-BPS kinematics renders the above amplitude two important features. Firstly, the amplitude implicitly depends on a special $SL(2,\mathbb{C})$ spinor $|u]$ (or equivalently the spinor $|u\rangle$) via its dependence on the reference spinor. Secondly, this dependence also means that the amplitude is written in a special little group frame where each momentum is decomposed along the directions $|u], |u\rangle$ and orthogonal directions $|i^w], |i^w\rangle$ as emphasized in \cite{Herderschee:2019dmc}. We will return to these issues in the later sections.

\section{Three-point amplitudes and SpGr(3,6)}
\label{sec:3 three-point amplitude}
In this section, we first rewrite the three-point amplitude and observe clues of a symplectic Grassmannian description. We will also discuss how the geometry of the massless amplitude is encoded in this rewriting of the massive amplitude. We will then understand the three-point special BPS kinematics in terms of the geometry of the symplectic Grassmannian. Concretely, we show how the symplectic Grassmannian geometry arises from momentum and mass conservation, leading to a geometric interpretation of the special kinematics $u$-variables.  This will prepare us for the next section where we will provide a symplectic Grassmannian integral formula for the three-point amplitude. 

\subsection{Clues from the three-point amplitude}
The overall kinematic factor in \eqref{3-point-amp} can be manipulated in the following manner
\begin{equation*}
\begin{aligned}
	\langle q| p_1p_3|q\rangle &= \langle q 1^I\rangle [1_I3_K]\langle 3^K q\rangle \\
	&=\langle q 1^I\rangle \left(-\langle 1_I3_K\rangle + u_{1I}u_{3K}\right)\langle 3^K q\rangle\\
	&=-\langle q u\rangle ^2,
\end{aligned}    
\end{equation*}
where from the first to the second line we have used the definition of the $u_i^I$ variables and from the second to the third line we have used spin sums. We can also write,  
\begin{align*}
	[q|p_1p_3|q]=-[qu]^2.
\end{align*}
This allows us to rewrite the amplitude as
\begin{align*}
	A^{(3)}&=-\frac{1}{\langle q u \rangle^2 [qu]^2}\delta^{(2)}([qQ])\delta^{(2)}([uQ])\delta^{(2)}(\langle qQ^\dagger\rangle),
\end{align*}
where the two delta functions in the middle can be written either in terms of square or angle brackets, since $\delta^{(2)}([uQ])=\delta^{(2)}(\langle uQ^\dagger\rangle)$. We can also choose the reference spinors such that $\langle qu\rangle = 1 =[qu]$. The amplitude is then written as
\begin{align*}
	A^{(3)}&=-\delta^{(2)}([qQ])\delta^{(2)}([uQ])\delta^{(2)}(\langle qQ^\dagger\rangle).
\end{align*}
We will ignore the overall minus sign in the following discussions, as it is a matter of normalization. Now, to see signs of the symplectic Grassmannian in the amplitude, we can write the above amplitude as
\begin{align}\label{3-point-Cstar}
	\mathcal{A}_3 = \delta^6\left(C_*.\Omega.\eta^T\right),
\end{align}
where
\begin{align} \label{C for three point}
 C_*&= \left(\begin{matrix}
	\langle q1^1\rangle & \langle q2^1\rangle & \langle q3^1\rangle \\
	[u1^1] & [u2^1] & [u3^1] \\
	[q1^1] & [q2^1] & [q3^1]
\end{matrix} \ \ \ \ \begin{matrix}\langle q1^2\rangle & \langle q2^2\rangle & \langle q3^2\rangle \\
	[u1^2] & [u2^2] & [u3^2] \\
	[q1^2] & [q2^2] & [q3^2]
\end{matrix} \right) ~, 
\end{align}
and
\begin{align*}
\eta &= \left(\begin{matrix}
	\eta_1^1& \eta_2^1 & \eta_3^1 & \ \ & \eta_1^2 & \eta_2^2 & \eta_3^2 \\
\end{matrix}\right), \\
\Omega&=\begin{pmatrix}
	0_{3\times 3} & \mathbb{I}_{3\times 3}\\
	-\mathbb{I}_{3\times 3} & 0_{3\times 3}
\end{pmatrix}.
\end{align*}
There are several features one can notice from this way of packaging the amplitude. First, one can easily check that due to momentum and mass conservation, the $C_*$ matrix above satisfies symplectic conditions, namely:
\begin{equation}\label{symp-cond}
C_*.\Omega. C_*^T = 0, \quad  C_*.\Omega.\Lambda^T = 0, \quad \text{and} \quad C_*.\Omega.\tilde{\Lambda}^T = 0,
\end{equation}
where,
\begin{equation}
    \begin{aligned}
	\Lambda &= \left(\begin{matrix}
		|1^1] & 	|2^1]  & |3^1] & \ \ & |1^2] & 	|2^2]  & |3^2] \\
	\end{matrix}\right),\\
	\tilde{\Lambda} &= \left(\begin{matrix}
		|1^1\rangle & |2^1\rangle & |3^1\rangle & \ \ & |1^2\rangle & |2^2\rangle & |3^2\rangle \\
	\end{matrix}\right).
\end{aligned}
\end{equation}
The conditions in \eqref{symp-cond} hint towards a symplectic Grassmannian $\SpGr(3,6)$ description of the amplitude. These conditions are similar to the ones found in the context of six dimensional rational maps in \cite{Cachazo:2018hqa} and used in \cite{Bering:2022tdr} to construct a six dimensional symplectic Grassmannian integral. The representation in \eqref{3-point-Cstar} also makes it clear that for any $A \in \SL(3,\mathbb{C})$, the matrix $AC_*$ would lead to the same amplitude. Thus we see an $\SL(3,\mathbb{C})$ invariance in this form of the amplitude that is not obvious from the kinematic space where the Lorentz group is $\SL(2,\mathbb{C})\times \SL(2,\mathbb{C})$. In fact, to uplift the matrix $C_*$ satisfying the symplectic conditions above to a point in a symplectic Grassmannian we need to produce the $\GL(3,\mathbb{C})$ freedom. While it is clear from the above that the amplitude \emph{shouts out} a $\SpGr(3,6)$ geometry, in the next subsection we will understand how such a description emerges by investigating the kinematic space. Before going into this detailed analysis, let us understand how to see the massless limit from this way of packaging the amplitude. 

\paragraph{Massless limit:} In the massless limit, the three-point Coulomb branch amplitude in \eqref{3-point-amp} reproduces the MHV and anti-MHV three-point amplitudes at the origin of the moduli space, as discussed in \cite{Herderschee:2019dmc}. Let us try to understand what happens to the matrix $C_*$ under this limit. This will give us hints on embedding the Grassmannians $\text{Gr}(1,3)$ and $\text{Gr}(2,3)$, describing the massless three-point amplitudes, inside $\SpGr(3,6)$. Recall that in the massless limit we have:
\begin{align*}
	|i^1\rangle &\to 0 ~~, \qquad |i^2\rangle \to -|i\rangle ~,\\
	|i^1] &\to |i] ~~, \qquad |i^2] \to 0~.
\end{align*}
Now, recall the special three body kinematics:
\begin{align*}
	\langle i^Ij^K\rangle + [i^Ij^K] = u_i^{I}u_j^K ~.
\end{align*}
Analyzing particular components of this equation in the massless limit, we obtain:
\begin{align*}
	u_i^1\,u_j^1 = [ij] ~~, \qquad u_i^2\,u_j^2 = \langle ij\rangle  ~~, \qquad u_i^1 u_j^2 = 0~~.
\end{align*}
Since $u_i^1 u_j^2 = 0\ \  \forall\  i,j$, we either have  $u_i^1 = 0 \ \ \forall \ i $ or $u_i^2 = 0 \ \ \forall \ i$. Note that other possible solutions result in restricted kinematics, and we discard such solutions. 

Let us consider the case of $u_i^2 = 0 \ \ \forall \ i~.$ Consequently, we have
\begin{align*}
	|u\rangle = u_{iI}|i^I\rangle \to u_{i}^1|i^2\rangle = -u_i^1|i\rangle ~~, \qquad |u] \to 0~.
\end{align*}
All the angle brackets are then proportional to each other. We can see this from $u_i^2\,u_j^2 = \langle ij\rangle = 0$ as well. We can find the proportionality factors as follows:
\begin{align*}
	&u_1^1 = \sqrt{\frac{(u_3^1 u_1^1)(u_1^1 u_2^1)}{(u_2^1 u_3^1)}} = \sqrt{\frac{[31][12]}{[23]}}~, \\
	\Rightarrow \quad |u\rangle &= -|1\rangle\sqrt{\frac{[31][12]}{[23]}} = -|2\rangle\sqrt{\frac{[12][23]}{[31]}} = -|3\rangle \sqrt{\frac{[23][31]}{[12]}}~.
\end{align*}
One can check that the relation $[ui^I] = -\langle ui^I\rangle$ is still valid. 

The massive momenta can be written in the special little group dictated by $u$ and $w$~as:
\begin{equation}\label{momenta in u w variables}
	p_i = |i^I\rangle[i_I| \left(u_{iJ}\,w_i^J\right) = -|i^I\rangle \left([i_J|\,u_{i}^J\,w_{iI} + [i^J|\,u_{iI}\,w_{iJ}\right) = |i^w\rangle[u| - |u\rangle [i^w|~.
\end{equation}
In the massless limit, we either have $u_i^1 \to 0$ leading to $|u\rangle\to 0$ and $p_i = |i^w\rangle [u|~.$ In this case, all the square spinors are equal to each other. Effectively (in the definition of $u$ variables), we have fixed the massless little group so that the square spinors are exactly equal to each other rather than just being proportional. Similarly, the other possible scenario is that $u_i^2 \to 0$ leading to $|u]\to 0$ and $p_i = -|i^w] \langle u|~.$ Then, all the angle spinors are equal to each other, and \eqref{momenta in u w variables} naturally clubs together the two massless special kinematics. 

Let us analyze the massless limit of $C_*$, as expressed in \eqref{C for three point}. Focusing on the first row, we have $\langle qi^1\rangle \to 0$, and
\begin{equation*}
	\langle q 1^2\rangle \to  -\langle q1\rangle = \langle qu\rangle\sqrt{\frac{[23]}{[31][12]}} = \sqrt{\frac{[23]}{[31][12]}} ~.
\end{equation*}
Similarly, we solve for $\langle q2^2\rangle$ and $\langle q3^2\rangle$. We obtain the following for the top row of the $C$:
\begin{equation*}
	\left( \begin{matrix}
		\langle qi^1\rangle & \langle qi^2\rangle 
	\end{matrix} \right) \to \frac{1}{\sqrt{[12][23][31]}}\left(\begin{matrix}
		0 & 0 & 0 & \  & [23] & [31] & [12]
	\end{matrix}\right)
\end{equation*}
To handle the bottom two rows, let us consider the following $\SL(3)$ transformation:
\begin{equation*}
	T = \left( \begin{matrix}
		1 & 0_{1\times 2}  \\ 
		0_{2\times 1} & \left(\begin{matrix}
			[q|^1 & [q|^2 \\
			[u|^1 & [u|^2
		\end{matrix}\right)
	\end{matrix} \right)
\end{equation*}
Note that $\det T = [qu]=1$. Recall that the amplitude is invariant under such an $\SL(3,\mathbb{C})$ transformation on the matrix $C_*$. Therefore, equivalent to $C_*$, we have $C'_*$:
\begin{equation*}
	C'_* = T^{-1}.C_* = \left(\begin{matrix}
		\langle qi^1\rangle & \langle qi^2\rangle \\
		|i^1] & |i^2]
	\end{matrix}\right)~.
\end{equation*}
The matrix $C'_*$ has a nice massless limit:
\begin{equation*}
	C'_* \to \left(\begin{matrix}
		0 & 0 & 0 & \ & \frac{[23]}{\sqrt{[12][23][31]}}& \frac{[31]}{\sqrt{[12][23][31]}}& \frac{[12]}{\sqrt{[12][23][31]}} \\ 
		|1] & |2] & |3] & \ & 0 & 0 & 0
	\end{matrix}\right)~,
\end{equation*}
and the massless amplitude is $\delta^6\left(C'_*.\Omega.\eta^T\right)$, which is the anti-MHV amplitude written in the non-chiral $\eta$ variables. Similarly, if we had chosen $u_i^1 = 0 \ \ \forall \ i$, we would have obtained $[ij] = 0$ for all $i,j$. This would have led to the following:
\begin{equation*}
	C''_*  = \left(\begin{matrix}
		-|i^1\rangle & -|i^2\rangle \\ 
		[qi^1] & [qi^2]
	\end{matrix}\right) \to \left(\begin{matrix}
		0 & 0 & 0 & |1\rangle & |2\rangle & |3\rangle \\ 
		\frac{\langle 23\rangle}{\sqrt{\langle 12\rangle\langle 23\rangle\langle 31\rangle}} & \frac{\langle 31\rangle}{\sqrt{\langle 12\rangle\langle 23\rangle\langle 31\rangle}} &\frac{\langle 12\rangle}{\sqrt{\langle 12\rangle\langle 23\rangle\langle 31\rangle}} & 0 & 0 & 0
	\end{matrix}\right) 
\end{equation*}
The amplitude would be $\delta^6\left(C''_*.\Omega.\eta^T\right)$, which is the MHV three-point amplitude. 

The above discussion suggests an embedding of $\text{Gr}(1,3)$ or $\text{Gr}(2,3)$ inside $\SpGr(3,6)$~as,
\begin{equation*}
C_*=\left(\begin{matrix}
	0 & w_* \nonumber\\
	w_*^\perp & 0
\end{matrix}\right),
\end{equation*}
where $w_*$ belongs to either $\text{Gr}(1,3)$ or $\text{Gr}(2,3)$ and the null matrix is to be interpreted as $1 \times 3$ or $2 \times 3$ accordingly. Note that the form of the embedding is preserved under transformations of the type,
\begin{equation*}
	N = \left( \begin{matrix}
		t & 0_{1\times 2}  \\ 
		0_{2\times 1} & t^{-1}M_{2 \times 2}
	\end{matrix} \right)
\end{equation*}
with $M_{2\times 2}\in \SL(2,\mathbb{C})$ when $w_*$ is $1\times 3$ and transformations of the type 
\begin{equation*}
	N^\prime = \left( \begin{matrix}
		 t^{\prime -1}M^\prime_{2 \times 2} & 0_{2\times 1}  \\ 
		0_{1 \times 2} & t^\prime
	\end{matrix} \right),
\end{equation*}
with $M^\prime \in \SL(2,\mathbb{C})$ when $w_*$ is $2\times 3$. Both the cases above describe $\GL(2,\mathbb{C})$ subgroups of $\SL(3,\mathbb{C})$ under which our amplitude is invariant. Note that the $\GL(1)$ symmetry of the massless description is a subgroup of the $\SL(3,\mathbb{C})$ symmetry of the massive amplitude.

\subsection{Symplectic Grassmannian geometry from the kinematics}\label{symp-geom}
We will now try to make concrete the hints we obtained in the previous subsection by investigating the structure of the kinematic space of the Coulomb branch. To begin with, consider a $n-$point scattering on the Coulomb branch where the momenta are $p_1,p_2,\ldots,p_n$ and they satisfy momentum conservation. For the masses, we will consider the analytically continued conservation relation given in \eqref{analyt-mass-cons}. This kinematics can be encapsulated by considering matrices of spinor-helicity variable $\Lambda$ and $\tilde{\Lambda}$ defined as
\begin{align*}
    \Lambda &= \left(\begin{matrix}
        |1^1] & \cdots & |n^1] & \ \ & |1^2] & \cdots & |n^2] \\
    \end{matrix}\right)_{2\times 2n},\nonumber\\
   \tilde{\Lambda} &= \left(\begin{matrix}
        |1^1\rangle & \cdots & |n^1\rangle & \ \ & |1^2\rangle & \cdots & |n^2\rangle \\
    \end{matrix}\right)_{2\times 2n}.
\end{align*}
The rowspan of these matrices define $2-$planes in $2n$ dimensions, therefore they correspond to two points in the Grassmannian $\Gr(2,2n)$. In this language, analogous to the massless case, the spinor-helicity brackets are given by the Pl\"ucker coordinates. Further, the mass conservation relations can be written as
\begin{equation}\label{self-symplectic}
\begin{aligned}
\Lambda . \Omega .  \Lambda^T&=0,\\
\tilde{\Lambda} . \Omega . \tilde{\Lambda}^T&=0,    
\end{aligned}
\end{equation}
where $\Omega$ is the skew-symmetric matrix:
\begin{align*}
\Omega=\begin{pmatrix}
0_{n\times n} & \mathbb{I}_{n\times n}\\
-\mathbb{I}_{n\times n} & 0_{n\times n}
\end{pmatrix}.
\end{align*}
The symplectic conditions above imply that both $\Lambda$ and $\tilde{\Lambda}$ describe two symplectic $2$-planes, i.e., they are actually points in $\SpGr(2,2n)$. The momentum conservation condition imposes a relation between them, which reads:
\begin{equation}\label{mutual-symplectic}
\begin{aligned}
\Lambda . \Omega . \tilde{\Lambda}^T& = 0~.
\end{aligned}    
\end{equation}
Our goal in this section is to understand how the conditions above give rise to a special kinematic description for three points. To do this, let us understand the conditions above in terms of row spans of matrices $\Lambda$ and $\tilde{\Lambda}$. The condition \eqref{self-symplectic} says that if we consider any two $2n-$dimensional vectors $v_1, v_2 \in \text{rowspan}(\Lambda)$, then,
\begin{equation*}
\begin{aligned}\label{v-vt}
v_1 . \Omega . v_2^T=0.
\end{aligned}
\end{equation*}
Similarly, for any two $w_1, w_2 \in \text{rowspan}(\tilde{\Lambda})$, 
\begin{equation}\label{w-wt}
\begin{aligned}
w_1 . \Omega . w_2^T=0.
\end{aligned}
\end{equation}
The condition \eqref{mutual-symplectic} says that for any $v \in \text{rowspan}(\Lambda)$ and $w \in \text{rowspan}(\tilde{\Lambda})$,
\begin{align}\label{v-wt}
v . \Omega . w^T=0.
\end{align}
Let us now consider a space $\C$ defined as the direct sum of the row spans of $\Lambda$ and $\tilde{\Lambda}$. Naively, this would describe a $4$-plane in $2n$ dimensions. For generic $n$, this is true. However, $n=3$ is special by a dimension count as we will see below.

Begin by noting that, due to \eqref{v-vt}, \eqref{w-wt} and \eqref{v-wt}, the following is true:
\begin{align*}
a^T . \Omega . b=0, \quad \text{ for all } a,b \in \C.
\end{align*}
In mathematical terminology, the vector space $\C$ is $\Omega$-isotropic. The isotropic condition above imposes an inequality on the dimension of the space $\C$. Given a space $V \subset \mathbb{C}^{2n}$, we define $V^\perp$ as
\begin{align*}
V^\perp = \big\{v^\perp\ |\ v^T .\, \Omega \,.\, v^{\perp} = 0, \quad \text{for any } v\in V \big\}.
\end{align*}
Since $\Omega$ is a non-degenerate matrix it follows that
\begin{align*}
\dim(V) + \dim(V^\perp) = 2n.
\end{align*}
For our isotropic space we have $\C \subseteq \C^\perp$. Therefore, 
\begin{align*}
\dim(\C) \leq \dim(\C^\perp).
\end{align*}
Combined with the equality above, we obtain that for an isotropic space $\C \subset \mathbb{C}^{2n}$:
\begin{align*}
\dim(\C) \leq n.
\end{align*}
When $n=3$, the dimension of $\C$ is at most $3$ and for generic kinematics we will have
\begin{align*}
\dim(\C) = 3.
\end{align*}
Naively, we would have expected a four-dimensional space constructed out of row spans of $\Lambda$ and $\tilde{\Lambda}$. Therefore, this indicates that for three particles, the row spans of $\Lambda$ and $\tilde{\Lambda}$ have one-dimensional intersection space. Further, for general kinematics $\C^\perp = \C$ and therefore,
\begin{align}\label{C-omeg-CT}
C . \Omega . C^T=0,
\end{align}
where $C$ is a matrix whose rowspan is our isotropic space $\C$. 

Thus, the matrix $C$ represents a point in the symplectic Grassmannian $\SpGr(3,6)$ defined as a subvariety of the Grassmannian $\text{Gr}(3,6)$. Conversely, we can also argue that if we consider a matrix $C \in \SpGr(n,2n)$ that satisfies
\begin{equation}\label{C-Omega_Lambda-tilde-Lambda}
\begin{aligned}
C . \Omega . C^T & =0~,\\
C . \Omega . \Lambda^T &=0~,\\
C . \Omega . \tilde{\Lambda}^T &=0~,
\end{aligned}    
\end{equation}
then this implies conditions \eqref{self-symplectic} and \eqref{mutual-symplectic} on $\Lambda$ and $\tilde{\Lambda}$ and further, it imposes
\begin{equation}
{\rm rowspan}(\Lambda) \subseteq \C \quad \text{and} \quad {\rm rowspan}(\tilde{\Lambda}) \subseteq \C.
\end{equation}
For $n=3$ we have shown that as $\Lambda$ and $\tilde{\Lambda}$ have an intersection space of dimension one, their joint row span has dimensionality $3$. Thus for three points, any isotropic space $\C$ is uniquely (up to $\GL(3,\mathbb{C})$) represented by a matrix $C_{3\times 6}$ satisfying the above conditions. This correspondence continues to hold for $n=4$ as the expected dimension of the joint row space of $\Lambda$ and $\tilde{\Lambda}$ is $4$, which is the same as the dimension of the row span of a matrix $C_{4\times 8}$. We will come back to the $n=4$ case in the next section.

If we label the columns of $C_{3\times 6}$ as $1^1, 2^1, 3^1, 1^2, 2^2, 3^2$, then in terms of Pl\"ucker coordinates the condition \eqref{C-omeg-CT} becomes the defining equation of the symplectic Grassmannian $\SpGr(3,6)$ as a subvariety of $\Gr(3,6)$:
\begin{align}\label{symp-ideal}
\sum_i (Si^1i^2)=0, \quad \text{for any }  S \in \{1^1, 2^1, 3^1, 1^2, 2^2, 3^2\}.
\end{align}
When we write the symplectic Grassmannian integral in the next section, we will be interested in looking for little group invariant functions of Pl\"ucker coordinates. Further, we will need to interpret the meaning of such functions in the expression of the amplitude that comes out of the integral. With this in mind, let us provide a representation of $C$ by using the spaces $\Lambda$ and $\tilde{\Lambda}$ that manifests the little group covariance of the Pl\"ucker coordinates as well as the defining equation of the symplectic ideal. 

Noticing that the spaces corresponding to $\Lambda_{2\times 6}$ and $\tilde{\Lambda}_{2\times 6}$ have an intersection of dimension one implies that there exists an $\SL(2,\mathbb{C})$ transformation
\begin{align*}
M_{2\times 2}=\begin{pmatrix}
|u] \\
|q]
\end{pmatrix},\, {[qu]=1},
\end{align*}
that acts on $\Lambda_{2\times 6}$, and another $\SL(2,\mathbb{C})$ transformation
\begin{align*}
N_{2\times 2}=\begin{pmatrix}
|u\rangle \\
|q\rangle
\end{pmatrix},\, {\langle qu\rangle=1},
\end{align*}
that acts on $\tilde{\Lambda}_{2\times 6}$, such that
\begin{align}\label{u-from-sl2}
[ui^I]=-\langle ui^I\rangle\equiv \uu_i^I.
\end{align}
That is, we can make two $\SL(2,\mathbb{C})$ transformations to align $\Lambda$ and $\tilde{\Lambda}$ such that one row of each matrix lies along the intersection space. Note that the $\SL(2,\mathbb{C})$ matrices above are not unique. We can consider residual $\SL(2,\mathbb{C})$ transformations that perform the scaling
\begin{align*}
    |u] &\rightarrow t|u],\, \qquad  |q] \rightarrow t^{-1}|q],\\
    |u\rangle &\rightarrow t|u\rangle,\, \qquad |q\rangle \rightarrow t^{-1}|q\rangle,
\end{align*}
which would still result in one row of realigned $\Lambda$ and $\tilde{\Lambda}$
to be along the intersection space. Therefore, our analysis above fixes $|u], |u\rangle$, and in turn $\uu_i^I$, only up to a global scalar factor that is independent of $i$. To consider invariants under the little group, let us further parametrize the $\Lambda_{2\times 6}$ matrix as
\begin{align*}
\Lambda = \begin{pmatrix}
m^\prime_1 \ww_1^1 \,& m^\prime_2 \ww_2^1 \,& m^\prime_3 \ww_3^1 \,& m^\prime_1 \ww_1^2 \,& m^\prime_2 \ww_2^2 \,& m^\prime_3 \ww_3^2\\
\uu_1^1 & \uu_2^1 & \uu_3^1 & \uu_1^2 & \uu_2^2 & \uu_3^2
\end{pmatrix},
\end{align*}
with $\ww_i^I\uu_{iI}=1$. The condition $\Lambda.\Omega.\Lambda^T$ is satisfied for the above, provided
\begin{align}\label{prime-m-cons}
\sum_i m^\prime_i=0.
\end{align}
Let us parametrize $\tilde{\Lambda}_{2\times 6}$ as,
\begin{equation*}
\resizebox{0.9\textwidth}{!}{$
\tilde{\Lambda} = \begin{pmatrix}
\tilde{m}_1 \ww_1^1 +b_1 \uu_1^1 \,& \tilde{m}_2 \ww_2^1 +b_2 \uu_2^1 \,& \tilde{m}_3 \ww_3^1 +b_3 \uu_3^1\,& \tilde{m}_1 \ww_1^2 +b_1 \uu_1^2 \,& \tilde{m}_2\ww_1^2 +b_2 \uu_2^2 \,& \tilde{m}_3 \ww_3^2 +b_3 \uu_3^2\\
\uu_1^1 & \uu_2^1 & \uu_3^1 & \uu_1^2 & \uu_2^2 & \uu_3^2
\end{pmatrix}$},
\end{equation*}
where we have used the same $\ww_i^I$ variables as before. The new degrees of freedom are in the $\tilde{m}_i, b_i$ variables. The condition $\tilde{\Lambda}.\Omega.\tilde{\Lambda}^T=0$ is satisfied provided,
\begin{align}\label{tilde-m-cons}
\sum_i \tilde{m}_i=0.
\end{align}
The condition $\Lambda.\Omega.\tilde{\Lambda}^T=0$ is satisfied provided,
\begin{align}\label{b-tildem-cons}
\sum_i b_i\tilde{m}_i=0.
\end{align}
Thus we can consider a $3\times 6$ matrix,
\begin{equation*}
\resizebox{0.9\textwidth}{!}{$
C_* = \begin{pmatrix}
m^\prime_1 \ww_1^1 \,& m^\prime_2 \ww_2^1 \,& m^\prime_3 \ww_3^1 \,& m^\prime_1 \ww_1^2 \,& m^\prime_2 \ww_2^2 \,& m^\prime_3 \ww_3^2\\
\uu_1^1 & \uu_2^1 & \uu_3^1 & \uu_1^2 & \uu_2^2 & \uu_3^2\\
\tilde{m}_1 \ww_1^1 +b_1 \uu_1^1 \,& \tilde{m}_2 \ww_2^1 +b_2 \uu_2^1 \,& \tilde{m}_3 \ww_3^1 +b_3 \uu_3^1\,& \tilde{m}_1 \ww_1^2 +b_1 \uu_1^2 \,& \tilde{m}_2 \ww_2^2 +b_2 \uu_2^2 \,& \tilde{m}_3 \ww_3^2 +b_3 \uu_3^2
\end{pmatrix}$},
\end{equation*}
that satisfies \eqref{C-omeg-CT} and \eqref{C-Omega_Lambda-tilde-Lambda} provided the parameters  above\footnote{When $m_i^\prime=\tilde{m}_i$, then the above $\uu_{iI}, \ww_{iI}$ variables can be rewritten in terms of the $u_{iI}, w_{iI}$ of~\cite{Herderschee:2019dmc}
\begin{align*}
    \uu_i^I&=m_i u_i^I\\
    \ww_{iI}&=\frac{1}{m_i}w_{iI}.
\end{align*}
} satisfy $\uu_i^I\ww_{iI}=1$ and equations \eqref{prime-m-cons}, \eqref{tilde-m-cons}, \eqref{b-tildem-cons}.
We can now compute the Pl\"ucker coordinates in terms of this parametrization.
\begin{align*}
(i^Ij^Jk^K)=(m^\prime_i\tilde{m}_j-m^\prime_j\tilde{m}_i) \ww_i^I\ww_j^J\uu_k^K+\tilde{m}_i(b_j-b_k)\ww_i^I\uu_j^J\uu_k^K+\text{cyclic},
\end{align*}
where the cyclic sum is over the cyclic interchange of the pairs of labels $\{i,I\}, \{j,J\}, \{k,K\}$. For the particular case where we have $(i^1i^2j^J)$ we obtain,
\begin{align*}
(i^1i^2j^J)=-\tilde{m}_i(b_i-b_j)\uu_j^J+(m^\prime_i\tilde{m}_j-m^\prime_j\tilde{m}_i)\ww_j^J.
\end{align*}
Due to \eqref{prime-m-cons}, \eqref{tilde-m-cons} and \eqref{b-tildem-cons} we can show that,
\begin{align*}
(i^1i^2j^J)=-(k^1k^2j^J),\, k\neq \{i,j\}.
\end{align*}
This is equivalent to the symplectic ideal condition \eqref{symp-ideal} as promised. From here, one can show that all the little group invariant combinations of the Pl\"ucker coordinates are zero. For instance, it follows from the symplectic ideal relation that
\begin{align*}
	(1^11^22^J)(3^13^22_J)=0.
\end{align*}
 One can also show that more non trivial combinations are zero due to the Pl\"ucker relations. For instance, the Pl\"ucker relations dictate that
\begin{align*}
(1^I2^J3^K)(1^L2_J3_K)\propto (2^J2_J1^I)(1^L3^K3_K).
\end{align*}
Then, the symplectic ideal condition implies the vanishing of the following little group invariant object
\begin{align*}
(2^12^21_I)(1^I2^J3^K)(1^L2_J3_K)(1_L2^12^2)=0.
\end{align*}
We can approach the general problem as follows. In a given algebraic variety, Derksen's algorithm \cite{derksen1999computation} provides a method to compute the generators of an invariant ring for a given linear reductive action $G$ via a Gr\"obner basis computation. We considered the invariant ring under the right action of $\SL(2,\mathbb{C})\times \SL(2,\mathbb{C})\times \SL(2,\mathbb{C})$ for the Grassmannian $\Gr(3,6)$. Running Derksen's algorithm on {\tt Macaulay2} \cite{M2}, we found a single generator for this invariant ring. This generator vanishes under the symplectic ideal! Thus, there are no little group invariant polynomials of Pl\"ucker coordinates in $\SpGr(3,6)$. Therefore, we will be able to write the symplectic Grassmannian integral in a special little group frame. There is a further subtlety in the fact that the kinematic space mapping to $\SpGr(3,6)$ we presented above was under the general case when $\tilde{m}_i\neq m^\prime_i$, whereas amplitudes in \cite{Herderschee:2019dmc} are formulated with $\tilde{m}_i=m^\prime_i$. The discussions in this section illustrate that we can formulate the symplectic Grassmannian description in terms of special kinematic $u$-- and $w$-- variables even in the general case. For the rest of this manuscript, we consider $\tilde{m}_i=m^\prime_i$ to guide us in writing the integrand for the symplectic Grassmannian integral.

\section{Symplectic Grassmannian integrals} \label{sec:4 symplectic grassmannian integrals}

In this section, we write an integral over the symplectic Grassmannian that evaluates to give us the Coulomb branch amplitudes. We comment on the general properties expected from this integral. Afterward, we explicitly evaluate the integral for the three and four-particle cases and obtain the Coulomb branch amplitude. 

We have an integral over $C$, an $n$-plane in a $ 2n$-dimensional space. The following are the features of the integral:
\begin{itemize}
    \item The symplectic Grassmannian space $C$ is symplectically orthogonal to itself by definition. $C.\Omega.C^T$ is an $n\times n$ antisymmetric matrix, so we include the following delta function in the integral: $\delta^{n(n-1)/2}\left(C.\Omega.C^T\right)~.$ 
    \item The momentum conservation is written as $\Lambda .\Omega . \tilde{\Lambda}^T =0$, so the $\Lambda$ and $\tilde{\Lambda}$ are symplectically orthogonal to each other. Also, the central charge (mass) conservation, $\Lambda.\Omega.\Lambda^T=0=\tilde{\Lambda}.\Omega.\tilde{\Lambda}^T$, enforces that the $\Lambda$ and $\tilde{\Lambda}$ planes are symplectically orthogonal to themselves as well. We wish to linearize these constraints using the Grassmannian. So, we propose that the delta functions in the integral constraints the $C$ to be orthogonal to both $\Lambda$ and $\tilde{\Lambda}$: $\delta^{2n}\left(C.\Omega.\Lambda^T\right)\,\delta^{2n}\left(C.\Omega.\tilde{\Lambda}^T\right)$. 
    \item The anticommuting Grassmann variables $\{\eta^a_{iI}\}$ form a $(2\times 2n)$ space, and we demand that there be $\delta^{2n}\left(C.\Omega.\eta^T\right)$ in the integral. This should produce the appropriate supercharges. 
    \item The integral has the function $f_n\left(\{M_i\}\right)$. It should be cyclically symmetrical. Due to the overall $\SL(n)$ invariance of the integral, $f_n$ is written in terms of the minors of $C$: $\{M_i\}$. For the integral to have a complete GL$(n)$ invariance, $f_n$ should have a weight equivalent to $(1-n)$ minors of $C$. One can check that upon $C \to G C$, $G$ being an arbitrary $\GL(n)$ element, the integral scales as follows:

\begin{equation}\label{scaling of C minors} 
\resizebox{0.9\textwidth}{!}{$
\begin{aligned}
    &\bigg\{\dd^{2n^2}C\ \delta^{n(n-1)/2}\big(C.\Omega.C^T\big)\,\delta^{2n}\big(C.\Omega.\Lambda^T\big)\,\delta^{2n}\big(C.\Omega.\tilde{\Lambda}^T\big)\,\delta^{2n}\big(C.\Omega.\eta^T\big)\bigg\} \nonumber \\
    & \to |\text{det } G|^{n-1} \times \bigg\{\dd^{2n^2}C\ \delta^{n(n-1)/2}\big(C.\Omega.C^T\big)\,\delta^{2n}\big(C.\Omega.\Lambda^T\big)\,\delta^{2n}\big(C.\Omega.\tilde{\Lambda}^T\big)\,\delta^{2n}\big(C.\Omega.\eta^T\big)\bigg\}
\end{aligned}$}
\end{equation}

Note that the general form of $f_n$ is not known, and we speculate that it may depend on the external kinematics $\Lambda$ and $\tilde{\Lambda}$ for the general cases, as it does for the gravitational theories \cite{Armstrong:2020ljm,Heslop:2016plj}. This was also argued in \cite{Bering:2022tdr} for the six-dimensional $(1,1)$ theory.
\end{itemize}
Putting all these together, we have the following integral:
\begin{equation} \label{actual integral}
\resizebox{0.92\textwidth}{!}{$
\begin{aligned}
    \Aboxed{ \ \ \ F_n = \int \frac{\dd ^{2n^2}C}{\text{GL}(n)}\ \delta^{n(n-1)/2}\big(C.\Omega.C^T\big) \,f_n\big(\{M_i\}\big)\,\delta^{2n}\big(C.\Omega.\Lambda^T\big) \ \delta^{2n}\big(C.\Omega.\tilde{\Lambda}^T\big) \ \delta^{2n}\big(C.\Omega.\eta^T\big)  \ \ \ } 
\end{aligned}
$}
\end{equation}

One of the points worth noting is the fact that $C$ is symplectically orthogonal to itself, but also to $\Lambda$ and $\tilde{\Lambda}$. This implies that \textbf{C contains both the $\Lambda$ and $\tilde{\Lambda}$ planes in it.} As we in the earlier section, this leads to special kinematics for three particles, as $C$ does not have enough degrees of freedom to encode independent $\Lambda$ and $\tilde{\Lambda}$, leading to constraints on $\Lambda$ and $\tilde{\Lambda}$. 

After performing the integration over $C$, we obtain the momentum, mass, and supercharge conservation. There are a total of $2n^2 - n^2 - n(n-1)/2$ degrees of freedom in $C$, which are being fixed by $4n$ delta functions. We expect six overall delta functions constraining the external kinematics out of the total $4n$: $\sum |i^I\rangle [i_I| = 0~,$ $\sum \langle i^1i^2\rangle=0~,$ $\sum [i^1i^2]=0$. Note that we have analytically continued the central charge (mass) $m$ into the complex plane. This is natural if we interpret the particular Coulomb branch theory as the dimensional reduction of a six-dimensional theory, leading us to six `momentum conservation' conditions on the external kinematics. We shall study this in more detail in section \ref{sec:5 six dimensional theory}. 

One can morph the available delta functions into overall momentum and mass conservation. This leaves us with a total of $n(n+1)/2 - (4n-6)$ delta functions. For $n=3$ and $n=4$, this number evaluates up to zero. The $C$ is completely fixed using the available delta functions. Like in the massless case, for higher $n$, the rest of the integrals pick up the appropriate residues of the $f_n$. In appendix \ref{Appendix:D evaluation of the grassmannian integrals}, we show that for $n=3$ and $n=4$, the following identity holds:
\begin{align}
    &\int \frac{\dd^{n\times 2n}C}{\GL(n)}\,\delta^{n(n-1)/2}\big(C.\Omega.C^T\big)\,\delta^{2n}\,\big(C.\Omega.\Lambda^T\big)\,\delta^{2n}\big(C.\Omega.\tilde{\Lambda}^T\big) \nonumber \\
    &\hspace{4cm}= \delta\left(\sum \langle i^1i^2\rangle\right)\,\delta\left(\sum [ i^1i^2]\right)\,\delta^4\left(\sum |i^I\rangle [i_I|\right)\ \bigg|_{C = C_*} \label{carrying our the integral}
\end{align}

We have obtained two mass conservation delta functions $\sum \langle i^1i^2\rangle =0 $ and $\sum [i^1i^2]=0~.$ Therefore in the description where $\langle i^1i^2\rangle = - [i^1i^2]~.$, we have a $\delta(0)$ tagging along. The symplectic Grassmannian is a natural description for four-dimensional amplitudes with complex mass, such that $\langle i^1i^2\rangle$ and $[i^1i^2]$ are independent of each other. This is equivalent to five on-shell kinematic variables for each momentum, which is the same for massless six-dimensional momentum. We are getting a six-dimensional momentum conservation condition due to the symplectic Grassmannian. We have already studied this complexified four-dimensional kinematics in the previous section. In section \ref{sec:5 six dimensional theory}, we shall see the connection of six-dimensional SYM amplitudes with the symplectic Grassmannian. 

As we mentioned earlier, we will use the formulation of the amplitude with $\langle i^1i^2\rangle = - [i^1i^2]~$ as a guideline to construct the symplectic Grassmannian integral. Strictly speaking, this would imply the appearance of a $\delta(0)$ in the amplitude. However, from our discussion in the previous section, the analysis used to construct the integral can easily be generalised to the case without this condition. The appearance of the $\delta(0)$ when imposing this additional condition can be physically interpreted as the $\delta(0)$ one would obtain upon dimensionally reducing the amplitude from six to five dimensions. As an analogy, consider a scattering process in four dimensions with all the momenta in a single two-dimensional (spatial) plane, so that $p_i^3 = 0$ for all momenta. In such a case, $\delta^4(\sum p_i)$ contains a $\delta(0)$ corresponding to the $\delta(\sum p_i^3)$~. Coulomb branch amplitudes with four kinematics degrees of freedom for each momentum correspond to a scattering in a five-dimensional hyperplane of the six-dimensional space. So, the $\delta(0)$ is a consequence of the compactification of $\mathcal{A}\ \delta^6(\sum p_i)$ such that no momentum is allowed in a particular compact direction. The symplectic Grassmannian is naturally suited for the six-dimensional maximal SYM theory. 

\subsection{Three particle amplitude}
Taking into account the Jacobian factors \eqref{carrying our the integral}, the integral in \eqref{actual integral} is evaluated to the following:
\begin{equation}\label{actual integral for 3 point}
\begin{aligned}
    F_3 &= \int \frac{\dd ^{3\times 6}C}{\text{GL}(3)}\ \delta^{3}\big(C.\Omega.C^T\big) \,f_3\big(\{M_i\}\big)\,\delta^{6}\big(C.\Omega.\Lambda^T\big) \ \delta^{6}\big(C.\Omega.\tilde{\Lambda}^T\big) \ \delta^{6}\big(C.\Omega.\eta^T\big) \\
    &= f_3\big(\{M_i\}\big)\bigg|_{C = C_*}\ \delta^{6}\big(C_*.\Omega.\eta^T\big) \delta\left(\sum \langle i^1i^2\rangle\right)\,\delta\left(\sum [ i^1i^2]\right)\,\delta^4\left(\sum p_i\right) 
\end{aligned}
\end{equation}
The delta functions localize the $C$ matrix for three particles as follows\footnote{Note that the solution below is when the spinor-helicity variables satisfy $\langle i^1i^2\rangle = -[i^1i^2]$. The solution can also be written down when this condition is not satisfied, as discussed in section-\ref{symp-geom} . However for fixing the integrand both of these forms are adequate. Firstly, the amplitude in \cite{Herderschee:2019dmc} was formulated by assuming these conditions. Secondly, facts about the symplectic Grassmannian such as there being no little group invariant combinations of minors is true in both the formulations. Therefore, we will consider this simpler solution to fix our integrand.}:
\begin{equation} \label{C matrix}
    C_* = \left( \begin{matrix}
        \ang{q 1^1} & \ang{q 2^1} & \ang{q 3^1} & & & & \ang{q 1^2} & \ang{q 2^2} & \ang{q 3^2} \\
        \ang{u 1^1} & \ang{u 2^1} & \ang{u 3^1} & & & & \ang{u 1^2} & \ang{u 2^2} & \ang{u 3^2} \\
        [q 1^1] & [q 2^1] & [q 3^1] & & & & [q 1^2] & [q 2^2] & [q 3^2] 
    \end{matrix} \right)   
\end{equation}
We refer the reader to Appendix \ref{Appendix:D evaluation of the grassmannian integrals} for details of the integral. 

The fermionic delta functions in the integral evaluate to give the correct supercharges as follows:
\begin{align}\label{supercharges for n=3} 
    \delta^{2\times 3}\left(C_*.\Omega.\eta^T\right) = \delta^2\left( \langle q| \sum |i^I\rangle \eta_{iI}\right) \,\delta^2\left( \langle u| \sum |i^I\rangle \eta_{iI}\right) \,\delta^2\left( [ u| \sum |i^I] \eta_{iI}\right)
\end{align}
The three-point Coulomb branch amplitude is precisely $\delta^6\big(C_*.\Omega.\eta^T\big)~.$ We infer that $F_3$ \eqref{actual integral for 3 point} corresponds to the three-point amplitude, as long as $f_3(\{M_i\})|_{C=C_*} = 1$~.

\vspace{0.5cm}

{\bf Fixing the integrand} \\
In the earlier analysis, we argued that the integrand has a (cyclically invariant) function $f_n$ made up of minors of $C$. Here we present multiple choices of $f_3$, which evaluate to the identity on the solution $C= C_*$~. For instance, one choice is as follows:
\begin{align}
    f_3 = \frac{1}{(1^u2^w3^w)(2^u3^w1^w)} + \frac{1}{(2^u3^w1^w)(3^u1^w2^w)} + \frac{1}{(3^u1^w2^w)(1^u2^w3^w)}~. \label{f_3 minor combinations}
\end{align}
Here, $(i^uj^wk^w)$ is a minor of $C$, written in a particular little group frame dictated by the special kinematics $\left(1=u_{iJ}\,w_i^J\right)$:
\begin{align}
    p_i = |i^I\rangle[i_I| \left(u_{iJ}\,w_i^J\right) = -|i^I\rangle \left([i_J|\,u_{i}^J\,w_{iI} + [i^J|\,u_{iI}\,w_{iJ}\right) = |i^w\rangle[u| - |u\rangle [i^w|~.
\end{align}
In these variables, we label the columns of $C$ as $\{1^w \ , \ 1^u \ , \ 2^w \ , \ 2^u \ ,  \ 3^w \ , \ 3^u\}~.$ The minors $(1^w2^u3^w)$, $(2^w3^u1^w)$ and $(3^w1^u2^w)$ appear in the $f_3$ \eqref{f_3 minor combinations}. 

The minors in the special little group frame are related to the minors of $C$ written in the generic little group as follows:
\begin{align*}
    (i^uj^wk^w) = u_{iI}\,w_{jJ}\,w_{kK}\,(i^Ij^Jk^K)~.
\end{align*}

Let us focus on the solution $C_*$, \eqref{C matrix}. For the solution, we can compute the minor $(i^Ij^Jk^K)$\footnote{To obtain the above relation we used $\langle i^I j^J\rangle + [i^Ij^J] = u_i^Iu_j^J$, which holds for any $(i,j)$ in $\{(1,2)\,,\,(2,3)\,,\,(3,1)\}~.$ So, we have $\langle 1^I2^J\rangle + [1^I2^J] = u_1^Iu_2^J$, and $\langle 2^J1^I\rangle + [2^J1^I] = -u_2^Ju_1^I~.$}:
\begin{subequations}
\begin{align}
    (1^I2^J3^K) &= \langle q1^I\rangle u_{2}^Ju_3^K + \langle q2^J\rangle u_{3}^Ku_1^I + \langle q3^K\rangle u_{1}^Iu_2^J~~,  \\
    (i^Ij^Ji^K) &= \left(\langle qi^K\rangle u_i^I - \langle qi^I\rangle u_i^K\right)u_j^J \hspace{2cm}  (i,j) \in \{(1,2)\,,\,(2,3)\,,\,(3,1)\}~,\\
    (i^Ij^Ji^K) &= -\left(\langle qi^K\rangle u_i^I - \langle qi^I\rangle u_i^K\right)u_j^J \hspace{1.6cm}  (i,j) \in \{(2,1)\,,\,(3,2)\,,\,(1,3)\}~.
\end{align}
\end{subequations}
We observe for the particular solution \eqref{C matrix}, {\bf all the minors in the special little group frame are simply either ones or zeros}:
\begin{subequations}
    \begin{align}
    &(1^u2^u3^u) = (1^w2^w3^w) = 0 ~~, \\
    & (1^u2^u3^w) = (1^u2^w3^u) = (1^w2^u3^u) = 0 ~~, \\
    & (1^u2^w3^w) = (1^w2^u3^w) = (1^w2^w3^u) = 1 ~~,\\
    \text{for } & \ \  (i,j) \in \{(1,2)\,,\,(2,3)\,,\,(3,1)\}: \ \qquad  
     (i^ui^wj^u) =  0 ~~~, \qquad (i^ui^wj^w) = 1 ~,\\
    \text{for } & \ \  (i,j) \in \{(2,1)\,,\,(3,2)\,,\,(1,3)\}: \ \qquad  
     (i^ui^wj^u) =  0 ~~~, \qquad (i^ui^wj^w) = -1~.  
\end{align}    
\end{subequations}
Note that the symplectic condition for the Plucker coordinates is $\sum_i (i^ui^wk^J) = 0~.$ This is naturally satisfied by the above relations.  

The minors' simple nature also manifests that { \bf all the little group invariant combinations of the minors must be zero. } Note that any little group invariant combination must have equal number of $(\bullet)^u$s and $(\bullet)^w$s, but the only non-zero minors are $(\bullet^u \bullet^w\bullet^w)$~. For instance,
\begin{align*}
    (1_I1^I2^J)(2_J3_K3^K) = 4\times (1^u1^w2^u)(2^w3^u3^w) - 4\times (1^u1^w2^w)(2^u3^u3^w) = 0~.
\end{align*}

We wish for the integral \eqref{actual integral} for $n=3$ to give the correct three-particle amplitude. Comparing the correct three-particle amplitude with the integral carried out in \eqref{n=3 integral carried out} and \eqref{supercharges for n=3}, we infer that $f_3$ should evaluate to identity at the particular solution for $C$ \eqref{C matrix}. For it to be cyclically invariant, the $f_3$ can be one of the following:
\begin{equation*}
    \begin{aligned}
    f_3 &= \frac{1}{(1^u2^w3^w)(2^u3^w1^w)} + \frac{1}{(2^u3^w1^w)(3^u1^w2^w)} + \frac{1}{(3^u1^w2^w)(1^u2^w3^w)} ~,\\
    \text{or } \ \ f_3 &= \frac{1}{(1^u2^w3^w)^2} + \frac{1}{(2^u3^w1^w)^2} + \frac{1}{(3^u1^w2^w)^2} ~,\\
    \text{or } \ \ f_3 &= \frac{1}{(1^u1^w2^w)^2} + \frac{1}{(2^u2^w3^w)^2} + \frac{1}{(3^u3^w2^w)^2} ~,\qquad  \text{etc.}  
\end{aligned}
\end{equation*}
All of these yield a cyclically invariant symplectic Grassmannian integral, evaluating up to the correct three-point amplitude. In order to choose one from the above, we need to determine what the `natural' integration measure is over the symplectic Grassmannian.

\subsection{Four particle amplitude}
The Coulomb branch four-particle amplitude is as follows:
\begin{align*}
    A_4 = \frac{1}{s_{12}s_{23}}\,\delta^4\left(|i^I\rangle
    \eta_{iI}\right)\,\delta^4\left(|i^I]
    \eta_{iI}\right)
\end{align*}
We have omitted the momentum and mass conservation delta functions. Note that $s_{ik}$ are the modified Mandelstam variables:\footnote{One can interpret these modified Mandelstam variables as Mandelstam variables in higher dimensions, imagining masses as momenta in compact directions.} $s_{ik} = - 2p_i.p_k - 2m_im_k~.$

From the arguments in section-\ref{symp-geom}, we know that for four points as well the dimensionality of the row spans of $\Lambda$ and $\tilde{\Lambda}$ add upto four which is appropriate for a $C_{4\times 8}$ that can parametrise $\SpGr(4,8)$. We can therefore construct the symplectic Grassmannian integral in this case by using this isomorphism as a guideline.

Let us focus on the integral \eqref{actual integral} for $n=4$. The delta functions localize the $C$ onto the following value:
\begin{align*}
    C_*{}_{4\times 8} = \left(\begin{matrix}\Lambda_{2\times 8} \\ 
    \tilde{\Lambda}_{2\times 8} \end{matrix}\right)~.
\end{align*}
Taking into account the appropriate Jacobian factors, we have \eqref{carrying our the integral}, we have:
\begin{equation}\label{actual integral for 4 point}
    \begin{aligned}
    F_4 &= \int \frac{\dd ^{4\times 8}C}{\text{GL}(4)}\ \delta^{6}\big(C.\Omega.C^T\big) \,f_4\big(\{M_i\}\big)\,\delta^{2\times 4}\big(C.\Omega.\Lambda^T\big) \ \delta^{2\times 4}\big(C.\Omega.\tilde{\Lambda}^T\big) \ \delta^{2\times 4}\big(C.\Omega.\eta^T\big) \\
    &= f_4\big(\{M_i\}\big)\big|_{C=C_*}\ \delta^{2\times 4}\left(C_*.\Omega.\eta^T\right)\,\delta\left(\sum \langle i^1i^2\rangle\right)\,\delta\left(\sum [ i^1i^2]\right)\,\delta^4\left(\sum p_i\right)
\end{aligned}
\end{equation}
The Grassmannian delta functions give the right supercharges:
\begin{align*}
    F_4 = f_4\big(\{M_i\}\big)\big|_{C=C_*}\ \delta^4\left(|i^I\rangle
    \eta_{iI}\right)\,\delta^4\left(|i^I]
    \eta_{iI}\right)\,\delta\left(\sum \langle i^1i^2\rangle\right)\,\delta\left(\sum [ i^1i^2]\right)\,\delta^4\left(\sum p_i\right)
\end{align*}

The maximum minors of this $C$ matrix are the modified Maldelstam variables:
\begin{align*}
    (i^1i^2k^1k^2) = -2p_i.p_k - 2m_i.m_k = s_{ik}~.
\end{align*}
As argued in \eqref{scaling of C minors}, $f_4$ must be cubic in the minors of $C$. A natural choice for $f_4$ is as follows:
\begin{align*}
    f_4 = \frac{1}{(1^11^22^12^2)(2^12^23^13^2)(3^13^24^14^2)} + \text{cyclic permutations}~.
\end{align*}
On the solution $C_*$, this particular $f_4$ evaluates to\footnote{Note that the modified Maldelstam variables satisfies $s_{12} + s_{23} + s_{13} =0~.$} :
\begin{align*}
    f_4\big|_{C = C_*} = \frac{2}{s_{12}^2s_{23}} + \frac{2}{s_{12}s_{23}^2} = \frac{-2s_{13}}{(s_{12}s_{23})^2} ~.
\end{align*}
So, we conclude that with this particular $f_4$, we have a cyclic symmetric symplectic Grassmannian integral $F_4$, which evaluates to
\begin{align*}
    F_4 = \frac{-2s_{13}}{s_{12}s_{23}}\,\mathcal{A}_4~.
\end{align*} 

Another perspective is that $f_n$ contains explicit information of external kinematics. In that case, we have: 
\begin{align*}
    f_4 =  \frac{-s_{12}s_{23}}{2s_{13}}\left(\frac{1}{(1^11^22^12^2)(2^12^23^13^2)(3^13^24^14^2)} + \text{cyclic permutations}\right)~.
\end{align*}
This yields us $F_4 = \mathcal{A}_4$.

In the massless theory, the four-point tree-level amplitude is equal to the maximal cut of a four-point box diagram. However, in the massive theory, the maximal cut of a four-point box diagram is only proportional to the tree-level amplitude \cite{MdAbhishek:2023nvg}. So if $F_4$ is to be obtained by the amalgamation of $F_3$, i.e., the three-point amplitudes, $F_4$ evaluates to be only proportional to the four-point amplitude. We hope to understand these factors better upon studying the amalgamation of the three-point symplectic Grassmannian integrals in the future.

\section{Amplitudes for six-dimensional maximally supersymmetric YM} 
\label{sec:5 six dimensional theory}
This section discusses the maximally supersymmetric Yang-Mills theory, $\mathcal{N}=(1,1)$ in six dimensions. The massless supermultiplet in this theory is the same as the 1/2-BPS (short) massive multiplet in the case of $\mathcal{N}=4$ four dimensions. Interpreting the mass in four dimensions as the momentum in the compact directions, we can study the six-dimensional theory in the variables appropriate for four dimensions. 

We wish not to lose any (kinematic) information going from six to four dimensions, hence we regard mass as a complex parameter, keeping five on-shell kinematic degrees of freedom intact for each particle:
\begin{align*}
    -(p^0)^2 + (p^1)^2 + (p^2)^2+ (p^3)^2+ (p^4)^2+ (p^5)^2 &= 0 \\
    \Longleftrightarrow \quad -(p^0)^2 + (p^1)^2 + (p^2)^2+ (p^3)^2 &= -mm^* = -(p^4)^2 -(p^5)^2
\end{align*}
Note that the physical mass of the particle in four dimensions is $\sqrt{((p^4)^2 +(p^5)^2)}$. However, keeping $m$ as a complex parameter will be useful: $m = -ip^4 - p^5$. In appendix \ref{Appendix:B dim reduction of spinor-helicity variables}, we present a detailed description of how the six-dimensional massless spinor-helicity variables due to \cite{Cheung:2009dc} reduce to the four-dimensional massive spinor-helicity variables due to \cite{Arkani-Hamed:2017jhn}. The spinor helicity variables in six dimensions $\lambsix$ and $\lambtsix$ decompose into two set of four-dimensional massive spinor helicity variables as follows \eqref{six dim spinor-helicity going to massive ones}:
\begin{align}
    \lambsix^A_a\to \left(\begin{matrix}
        |p_I]_\alpha \\ - |p_I\rangle^{\aldot}
    \end{matrix}\right) ~~, \qquad \quad \lambtsix_{A\adot} \to \left(\begin{matrix}
        [\overline{p}_I|^{\alpha} \\ -\langle \overline{p}_I|_{\aldot}
    \end{matrix}\right)~.
\end{align}
The two sets of massive spinor helicity variables are related to each other as follows \eqref{barred variables i nterms of unbarred}: 
\begin{align}
    |\bari^K] = |i^K]\sqrt{\frac{m_i}{m_i^*}} ~~~, \qquad |\bari^K\rangle = |i^K\rangle \sqrt{\frac{m_i^*}{m_i}} ~~.
\end{align}
Please refer to the Appendix \ref{Appendix:B dim reduction of spinor-helicity variables} for the full details.

Similarly, there are sixteen supercharges in the $\mathcal{N}=(1,1)$ algebra in six dimensions and $\mathcal{N}=4$ algebra in four dimensions. In appendix \ref{Appendix:C dim reduction of susy algebra}, we have presented precisely how the two are related to each other. 

The supermultiplet for $\mathcal{N}=(1,1)$ theory, tying together the sixteen components is as follows: \cite{Cachazo:2018hqa}
\begin{align*}
    \Phi = \phi^{1\dot{1}} + \eta_a\psi^{a \dot{1}} + \widetilde{\eta}_{\adot} \psi^{\adot 1} + \eta_a\widetilde{\eta}_{\adot} A^{a\adot} + \hdots (\eta)^2(\widetilde{\eta})^2 \phi^{2\dot{2}}~.
\end{align*}
In the following subsections, we discuss the signatures of symplectic geometry underlying the amplitudes in six dimensions.

\subsection{Four point amplitude}
As explained in appendix \ref{Appendix:C dim reduction of susy algebra}, the multiplicatively realized supercharges are: $\lambsix^A_a\,\eta^a$ and $\lambtsix_{A\adot}\,\eta^{\adot}$~. So, the amplitude is given neatly as follows: \cite{Dennen:2009vk}
\begin{align*}
    \mathcal{A}_4 = \frac{1}{s\,t}\delta^4\left(\sum_i\lambsix^A_{ia}\,\eta_i^a \right)\delta^4\left(\sum_i\lambtsix_{iA\adot}\,\widetilde{\eta}_i^{\adot} \right)
\end{align*}
One can check upon analyzing the appropriate coefficient that we have the correct four-gluon amplitude: $\langle 1_a2_b3_c4_d\rangle[1_{\adot}2_{\bdot}3_{\dot{c}}4_{\dot{d}}]/st$ from \cite{Cheung:2009dc}. Expressing this in terms of the four-dimensional massive spinor-helicity variables, we have:
\begin{align*}
    \mathcal{A}_4 = \frac{1}{st}\delta^2\left(|i^I\rangle\eta_i\right)\,\delta^2\left(|i^I]\eta_i\right)\ \delta^2\left(|\bari^I\rangle\etabar_i\right)\,\delta^2\left(|\bari^I]\etabar_i\right)
\end{align*}
Note that in the above expression, we refer to $s = 2\psix_1.\psix_2 = 2p_1.p_2+m_1m_2^*+m_1^*m_2$. Comparing it with the Coulomb branch four-point amplitudes, we make the observation that,
\begin{align*}
    \mathcal{A}_4 &= \frac{1}{(1^11^22^12^2)(2^12^23^13^2)}\delta^4\left(C_*.\Omega.\eta^T\right)\,\delta^4\left(\overline{C}_*.\Omega.\etabar^T\right)~~, \\
    C_{*4\times 8} =& \left\{\lambsix^A_{ia}\right\} =  \left( \begin{matrix}
        \left\{|i^I\rangle\right\} \\ \left\{ |i^I] \right\}
    \end{matrix} \right) ~~, \qquad \quad \overline{C}_{*4\times 8} = \left\{\lambtsix_{iA\adot}\right\} = \left(\begin{matrix}
        \left\{|\bari^I\rangle\right\} \\ \left\{|\bari^I]\right\}
    \end{matrix}\right)~.
\end{align*}
$(1^11^22^12^2)$ is a minor of $C_*$ matrix, which evaluates to be the Mandelstam variable $s = 2\psix_1.\psix_2~.$ Note that the equivalent minors of the matrix $C_*$ and $\overline{C}_*$ are equal, $(1^11^22^12^2) = (\overline{1}^1\overline{1}^2\overline{2}^1\overline{2}^2)~.$

We leave the tempting act of expressing the six-dimensional amplitude as an integral over the $\SpGr(n,2n)$ space for future work. Here, we merely make some comments regarding the same. The bottleneck is that $\overline{C}_*$ depends on the $C_*$ in a non-linear way. For example, for the case of $n=4$, look at the relation \eqref{lambd in terms of lamdtilde}. 

\subsection{Three-point amplitude}
For the three-point, let us consider the multiplicatively realized supersymmetric delta function, $\delta^4\left(\sum_i\lambsix^A_{ia}\,\eta_i^a \right)$. The four components of this delta function are not independent. In fact, we can understand the redundancy in terms of the $u$ variables and the intersection space, as in the earlier sections. 

Consider the $\lambsix$:
\begin{align*}
    \overset{\scriptscriptstyle{(6)}}{\Lambda} = \left(\begin{matrix}
        \lambsix_{(i=1)}^{A,(a=1)} & \lambsix_{(i=2)}^{A,(a=1)} & \lambsix_{(i=3)}^{A,(a=1)} & \ \ & \lambsix_{(i=1)}^{A,(a=2)} & \lambsix_{(i=2)}^{A,(a=2)} & \lambsix_{(i=3)}^{A,(a=2)}
    \end{matrix}\right)_{4\times 6}
\end{align*}
 Let us express it in terms of the massive spinors for four dimensions, according to \eqref{six dim spinor-helicity going to massive ones}. 
\begin{align*}
    \overset{\scriptscriptstyle{(6)}}{\Lambda} = \left(\begin{matrix}
        |1^1] & |2^1] & |3^1] & \ \ & |1^2] & |2^2] & |3^2] \\
        -|1^1\rangle & -|2^1\rangle & -|3^1\rangle & \ \ & -|1^2\rangle & -|2^2\rangle & -|3^1\rangle
    \end{matrix}\right)
\end{align*}
The three-body special kinematics results in the following condition: $\langle \tilde{u} i^I\rangle = - [\tilde{u}i^I]~.$ This tells us that there are only three independent rows in $\overset{\scriptscriptstyle{(6)}}{\Lambda}$:
\begin{align*}
    C_*\equiv  \left(\begin{matrix}
        \langle \tilde{q} i^I\rangle \\ \langle \tilde{u} i^I\rangle \\ [\tilde{q} i^I]
    \end{matrix}\right)_{3\times 6} ~.
\end{align*}
We have defined $C_*$ to be comprising of the linearly independent rows of $\overset{\scriptscriptstyle{(6)}}{\Lambda} ~.$ Note that $|\tilde{q}\rangle$ is defined by the relation $\langle \tilde{q}\tilde{u}\rangle =1 ~.$ This also establishes that out of the four supercharges $\delta^4\left(\sum_i\lambsix^A_{ia}\,\eta_i^a \right)$, the independent ones are $\delta^3\left(C_*.\Omega.\eta^T\right)~.$ A similar story holds for the $\lambtsix$ due to the fact $[u\bari^I] = - \langle u\bari^I\rangle~.$ The independent supersymmetric delta functions are given by $\delta^3\left(\overline{C}_*.\Omega.\etabar\right)$, where $\overline{C}_*$ is given by the following:
\begin{align*}
    \overline{C}_* = \left( \begin{matrix}
        \langle q\bari^I\rangle  \\ \langle u\bari^I\rangle \\ [q\bari^I]
    \end{matrix} \right)_{3\times 6} ~.
\end{align*}
Putting together the information collected, the three-point amplitude is written as follows:
\begin{align}
    \mathcal{A}_3 = \delta^3\left(C_*.\Omega.\eta^T\right)\,\delta^3\left(\overline{C}_*.\Omega.\etabar^T\right)~. \label{three-point amplitude for six dimensional theory}
\end{align}

We claim that the above three-point amplitude matches the one presented in \cite{Dennen:2009vk}, and gives the correct component amplitudes:
\begin{equation*}
\resizebox{0.9\textwidth}{!}{$
    \mathcal{A}_3 = \big(\eta_{1u}\eta_{2u}+\eta_{2u}\eta_{3u}+\eta_{3u}\eta_{1u} \big)\left(\sum\eta_{iw}\right) \times \left(\etabar_{1\tilde{u}}\etabar_{2\tilde{u}} + \etabar_{2\tilde{u}}\etabar_{3\tilde{u}} + \etabar_{3\tilde{u}}\etabar_{1\tilde{u}}\right)\left(\sum \etabar_{i\tilde{w}}\right) ~$},
\end{equation*}
where we have $\eta_{iw} = w_i^{I}\eta_{iI}$ , $\eta_{iu} = u_{i}^I\eta_{iI}$ , $\etabar_{i\tilde{u}} = \tilde{u}^{iI}\etabar_{iI}$ , and $\etabar_{i\tilde{w}} = \tilde{w}^{iI}\etabar_{iI}~.$
In appendix \ref{Appendix:E equivalent of three-point amplitude}, we show the equivalence. 

\section{Discussion}
\label{sec:6 discussion}
In this manuscript, we have studied the 3-point and 4-point Coulomb branch amplitudes in terms of the symplectic Grassmannian geometry. While the massive spinor-helicity variables themselves are related to $\SpGr(2,2n)\times \SpGr(2,2n)$, we showed that for 3-point and 4-point Coulomb branch kinematics, they are equivalent to a $\SpGr(n,2n)$. Further, we have also described the 3-point and 4-point amplitudes in terms of an integral over the symplectic Grassmannian. 

The three-point amplitudes in the Coulomb branch constructed in \cite{Herderschee:2019dmc} have several subtle complexities. Unlike the massless case, the super-amplitude requires the use of a reference spinor. This reference spinor depends implicitly on the three particle special BPS kinematics. In this work, we have given a geometric interpretation for the three particle special BPS kinematics by interpreting the $u$-spinors as the reference spinors in terms of the $\SL(2,\mathbb{C})\times \SL(2,\mathbb{C})$ that align the spinor-helicity $\Lambda$ and $\tilde{\Lambda}$ variables along their intersection space. However, the feature that the three-point super-amplitude is written in a special little group frame carries over to the symplectic Grassmannian description. We show this by noting that all the little group invariant combinations of the Pl\"ucker coordinates vanish under the symplectic conditions for the 3-point case. Thus, we are able to write a symplectic Grassmannian integral for the 3-point Coulomb branch amplitude in a special little group frame. 

For four point amplitudes, we are able to write the amplitude up to a known kinematic factor as the Grassmannian integral. Equivalently, one can say that the function of Pl\"cuker coordinates that multiplies the various delta functions in the symplectic Grassmannian integral also depends on (modified) Mandelstam variables. 

The Coulomb branch amplitudes of $\mathcal{N}=4$ SYM are related by dimensional uplift to the amplitudes of six dimensional $(1,1)$ SYM. We performed a dimensional reduction of the six dimensional spinor-helicity variables to four dimensional massive spinor-helicity variables in the little group covariant formalism. Using this dimensional reduction, we were able to write the six dimensional amplitudes in the four dimensional language where one can see two different matrices $C$ and $\bar{C}$ that satisfy symplectic relations. This elucidates the symplectic Grassmannian geometry of the six dimensional amplitudes in the four dimensional kinematic variables. However, framing the relationship between the two matrices above succinctly, which would allow a symplectic Grassmannian integral representation, is left for future work. 

In \cite{MdAbhishek:2023nvg}, the BCFW bridge construction was performed for the on-shell functions on the Coulomb branch. In this description, the four point tree level Coulomb branch amplitude is equivalent to a box on-shell diagram up to a known kinematic factor. It will be interesting to understand this in terms of amalgamations of the symplectic Grassmannian, analogous to the massless case. This may provide crucial insights into generalizing the symplectic Grassmannian integral description to higher points. In \cite{Herderschee:2019dmc}, tree amplitudes on the Coulomb branch were constructed up to five points. Writing a general expression for the $n$-point Coulomb branch amplitude is at least of the same complexity as writing a general expression for the $N^kMHV$ amplitudes in the massless case. Similar to the latter case, dual conformal invariance can provide a guideline to write such a higher point expression. We leave these questions for future work.

\section*{Acknowledgments}
We would like to thank Md. Abhishek, Nima Arkani-Hamed, Sujay Ashok, Jacob Bourjaily, Johannes Henn, Dileep P. Jatkar, Alok Laddha, Arkajyoti Manna, Julio Parra-Martinez, Arnab Priya Saha, Bernd Sturmfels, Jaroslav Trnka and Congkao Wen for discussions. We thank Sujay Ashok, Dileep P. Jatkar, Alok Laddha and Congkao Wen for detailed feedback on the manuscript. VCC and YE would like to thank the Max Planck Institute for Physics in Munich for hospitality. SH would like to thank Chennai Mathematical Institute, Indian Institute of Science Education and Research Pune, International Center for Theoretical Sciences (ICTS) Bengaluru, MPI for Mathematics in the Sciences Leipzig and Institut des Hautes Études Scientifiques for hospitality. AS would like to thank ICTS Bengaluru, the Indian Institute of Science (IISc) Bengaluru, the Institute for Advanced Study (IAS) Princeton, and Pennsylvania State University for hospitality, where part of this work was carried out. AS would like to thank Pennsylvania State University and the program ``Positive Geometry in Scattering Amplitudes and Cosmological Correlators" at ICTS, where a part of this work was presented. 

\medskip
VCC, YE and SH were funded by the European Union (ERC, UNIVERSE PLUS, 101118787). Views and opinions expressed are however those of the authors only and do not necessarily reflect those of the European Union or the European Research Council Executive Agency. Neither the European Union nor the granting authority can be held responsible for them.

\appendix

\section{Review of 1/2-BPS on-shell superspace}
\label{Appendix-Review-Onshell}
We will work with the little group covariant massive spinor-helicity formalism developed in \cite{Arkani-Hamed:2017jhn}. The momentum is expressed as a $2 \times 2$ matrix using the homomorphism between the complexified Lorentz group and $\SL(2,\mathbb{C})\times \SL(2,\mathbb{C})$ as reviewed in \cite{Elvang:2013cua},
	\begin{align*}
		p_{\alpha\dot{\beta}} = -\lambda_\alpha^I\tilde{\lambda}_{\dot{\beta}}^J\epsilon_{IJ},
	\end{align*}
	where $\alpha = 1,2$ and $\dot{\beta}=\dot{1},\dot{2}$ are the indices under the two $\SL(2,\mathbb{C})$'s. The matrix $\epsilon_{IJ}$ is given by,
	\begin{align*}
		\epsilon_{IJ}=\begin{bmatrix}
			0 & -1 \\
			1 & 0
		\end{bmatrix}.
	\end{align*}
    We will use the notation,
    \begin{align*}
    \lambda_\alpha^I&=|p]^I_\alpha,\qquad \tilde{\lambda}_{\dot{\beta}}^I=\langle p |^I_{\dot{\beta}},\nonumber\\
    \lambda^{I \alpha}&=[p|^{I\alpha},\qquad \tilde{\lambda}^{I\dot{\beta}}=| p \rangle^{I \dot{\beta}},
    \end{align*}
    where raising and lowering of $\SL(2,\mathbb{C})$ Lorentz group indices are carried out by using Levi-Civita symbols. Detailed conventions that we follow are given in \cite{MdAbhishek:2023nvg}. Here, we wish to highlight that the on-shell condition $p^2=-m^2$ is realized in the above formalism by using, 
    \begin{align*}
    \det(\lambda_\alpha^I)&=-[p^1p^2]=-\tilde{m},\nonumber\\
    \det(\tilde{\lambda}_{\dot{\beta}}^I)&=\langle p^1p^2\rangle=m^\prime,
    \end{align*}
    such that,
    \begin{align*}
    \tilde{m} \ m^\prime = m^2.
    \end{align*}
    We can also alternatively write,
    \begin{align}\label{spin-sums}
    |p_I]_\alpha [p^I|^\beta&=-|p^I]_\alpha [p_I|^\beta=\tilde{m}\delta^\beta_\alpha\nonumber\\
    |p^I\rangle \langle p_I|&=-|p_I\rangle \langle p^I|=m^\prime, 
    \end{align}
    which are known as the spin sums. Clearly, there is a $\GL(1)$ redundancy to rescale the parameters $\tilde{m}$ and $m^\prime$ such that $m^2$ remains the same. In the literature, one typically fixes this redundancy by setting
    \begin{align}\label{square-angle}
    \tilde{m} = m' = m.
    \end{align}
    When we write the amplitude as an integral over the Symplectic Grassmannian, we choose not to fix this redundancy. We will discuss it further in Sections \ref{sec:3 three-point amplitude} and \ref{sec:4 symplectic grassmannian integrals}.

    For the Coulomb branch of $\mathcal{N}=4$ SYM, there is an additional structure on the kinematic space due to supersymmetry \cite{Henn:2011xk,Craig:2011ws,Boels:2010mj, Herderschee:2019dmc}. We will use the little group covariant $1/2$-BPS on-shell superspace developed in \cite{Herderschee:2019dmc}. For a review of the spontaneous symmetry breaking and the $1/2$-BPS on-shell supermultiplet, see, for example, section 2 of \cite{Abhishek:2023lva}. In the description where one uses \eqref{square-angle}, the central charge corresponding to the Coulomb branch supermultiplet is,
    \begin{align*}
    Z_{AB} = \pm m \ \Omega_{AB},
    \end{align*}
    where $A,B=1,\ldots, 4$ are the R-symmetry indices and $\Omega_{AB}$ is the symplectic matrix with $\Omega^2=-I$. The positive and negative signs above are for BPS and anti-BPS multiplets respectively. From supersymmetry, it follows that the central charge needs to be conserved for the full amplitude. This translates to a condition on the masses,
    \begin{align*}
    \sum_{i=1}^n \pm m_i =0.
    \end{align*}
    A question we face in our work is how to generalize this for the case where we do not assume \eqref{square-angle}. We will take the kinematics on the Coulomb branch to be defined by the relations,
    \begin{align}\label{analyt-mass-cons}
    \sum_i \tilde{m}_i&=-\sum_i[i^1i^2]=0,\nonumber\\
    \sum_i m^\prime_i&=\sum_i \langle i^1i^2\rangle=0.
    \end{align}
    At the end, we approach the Coulomb branch kinematics used in the literature by using \eqref{square-angle} for BPS particles, and $\tilde{m}_i = m_i^\prime = -m_i$ for anti-BPS particles. 

    We also review the form of the super-charges here which we need in the main text. Similar to the six dimensional case \cite{Dennen:2009vk,Heydeman:2017yww}, it was found in \cite{Herderschee:2019dmc} that to write a little group covariant supermultiplet one needs to manifest only a $\SU(2)$ subgroup of the full R-symmetry group $\mathrm{USp}(4)$ on the Coulomb branch. This is realized by representing the $1/2$-BPS Coulomb branch multiplet in terms of a long multiplet with $N=2$ supersymmetry. As usual, half the supercharges are multiplicative and half are derivatives in the Grassmann variables $\eta_i^{Ia}$ where $a=1,2$ is the $\SU(2)$ R-symmetry index. The super-charges are given as,
    \begin{equation*}
    \begin{aligned}
    Q^{\dagger a \dot{\alpha}}&=\sum_i |i\rangle^{I\dot{\alpha}}\eta_{iI}^{a}~,\\
    Q_{a+2\,\alpha}&=\sum_i |i]^I_\alpha \eta_{iI}^a~,\\
    Q^{\dagger a+2 \,\dot{\alpha}}&=\sum_i |i\rangle^{I\dot{\alpha}}\frac{\partial}{\partial \eta_{iI}^{a}}~,\\
    Q_{a\alpha}&=\sum_i |i]^I_\alpha\frac{\partial}{\partial \eta_{iI}^{a}}~.
    \end{aligned}
    \end{equation*}
    One can then define the super-charge conserving delta functions as,
    \begin{equation*}
    \begin{aligned}
    \delta^{(4)}(Q)&=\prod_{a,\alpha}Q_{a+2\, \alpha}~~,\\
    \delta^{(4)}(Q^\dagger)&=\prod_{a,\alpha}Q^{\dagger a \dot{\alpha}}~~.
    \end{aligned}
    \end{equation*}
    For $n\ge 4$ multiplicity amplitudes, one has
    \begin{align*}
    \delta^{(4)}(Q)\delta^{(4)}(Q^\dagger)
    \end{align*}
    as a multiplicative factor in the scattering amplitude, ensuring half of super-charge conservation. The derivatively realized super-charges impose additional conditions which need to be solved to construct the super-amplitudes in a basis of Grassmann triads \cite{Herderschee:2019ofc}. The case of $3-$point amplitudes is special, we will review this in Section \ref{sec:3 three-point amplitude}.

\section{Dimensional reduction of six dimensional spinor-helicity variables} \label{Appendix:B dim reduction of spinor-helicity variables}
Here, we discuss the embedding of the massive spinor-helicity variables for four dimensions \cite{Arkani-Hamed:2017jhn} in the massless spinor-helicity variables for six dimensions \cite{Cheung:2009dc}. The goal is to describe all the five on-shell degrees of freedom for massless six-dimensional momenta $\psix^\mu$ in terms of the massive spinor-helicity variables of four dimensions.

\textbf{Conventions:}
We (roughly) follow the conventions set by Cheung and O’Connell in \cite{Cheung:2009dc} for the six-dimensional spinor-helicity variables and the usual ones for the four-dimensional variables.
\begin{itemize}
    \item The Lorentz group in six dimensions is $\SO(6) \cong \SU(4)$. The fundamental index of $\SU(4)$ is represented as $(\bullet)^{A,B,\hdots}$.
    \item The Lorentz group in four dimensions is $\SO(4) \cong \SU(2)\times \SU(2)$. The fundamental indices of the two $\SU(2)$s are represented as $(\bullet)^{\alpha,\beta,\hdots}$ and $(\bullet)^{\aldot,\bedot,\hdots}$.
    \item The little group for massless particles in six dimensions is $\SO(4) \cong \SU(2)\times \SU(2)$. The fundamental indices of the two $\SU(2)$s are represented as $(\bullet)^{a,b,\hdots}$ and $(\bullet)^{\adot,\bdot,\hdots}$.
    \item The little group for massive particles in four dimensions is $\SO(3) \cong \SU(2)$. The fundamental indices are represented as $(\bullet)^{I,J,\hdots}$.
    \item We represent the spinor-helicity variables and momenta in six dimensions with an explicit ${}^{\scriptscriptstyle{(6)}}$ over them: $\psix\,, \ \lambsix\,,\ \lambtsix.$ The variables without such a specification shall be four-dimensional by default. 
\end{itemize}

The spinor-helicity variables in six dimensions are defined as follows:
\begin{align}
    \psix^{AB} = \lambsix^{Aa}\, \lambsix^{Bb} \,\epsilon_{ab} ~, \qquad \qquad \psix_{AB} = \lambtsix_{A\adot} \, \lambtsix_{B \bdot}\,\epsilon^{\adot \bdot} \label{defining spinor-helicity for six dim}
\end{align}
A six-dimensional spinor decomposes into four-dimensional spinors in the following manner. 
\begin{align}\label{index decomposition}
    (\bullet)_{A = (1,2,3,4)} \to \left( \begin{matrix}
        (\bullet)^{\alpha = (1,2)} \\ (\bullet)_{\aldot = (\dot{1},\dot{2})}
    \end{matrix} \right)~~, \qquad \qquad     (\bullet)^{A = (1,2,3,4)} \to \left( \begin{matrix}
        (\bullet)_{\alpha = (1,2)} \\ (\bullet)^{\aldot = (\dot{1},\dot{2})}
    \end{matrix} \right) 
\end{align}
Note that the Lorentz group $\SU(4)$ broke down into $\SU(2)\times \SU(2)$ due to dimensional reduction. Similarly, the little group in six dimensions $\SU(2)\times \SU(2)$ breaks down into a single $\SU(2)$, the massive little group in four dimensions. So, after dimensional reduction, the indices $(\bullet)^{a,b\hdots}$ and $(\bullet)^{\adot,\bdot\hdots}$ are replaced by $(\bullet)^{I,J,\hdots}$.

In the particular representations of $\sigma$ matrices in six dimensions given in \cite{Cheung:2009dc}, we have:
\begin{align}
    \label{psix upper}
    \psix^{AB} = \left(\begin{matrix}
        \left(\begin{matrix}
            0 & -i p^4 + p^5 \\ 
            ip^4-p^5 & 0
        \end{matrix}\right)_{\alpha \beta} & p_{\alpha}{}^{\bedot} \\
        -p^{\aldot}{}_\beta & \left(\begin{matrix}
            0 & ip^4+p^5 \\ 
            -ip^4-p^5 & 0
        \end{matrix}\right)^{\aldot \bedot}
    \end{matrix}\right) \equiv \left(\begin{matrix}
         m^* \epsilon_{\alpha\beta} & p_\alpha{}^{\bedot} \\ -p^{\aldot}{}_\beta & -m \epsilon^{\aldot \bedot}
    \end{matrix}\right)
\end{align}
In the last equality, we have introduced a complex parameter $m = -ip^4-p^5$. Similarly, we have:
\begin{align*}
    \psix_{AB} = \left(\begin{matrix}
        -m\epsilon^{\alpha \beta} & p^\alpha{}_{\bedot} \\
        -p_{\aldot}{}^\beta & m^*\epsilon_{\aldot \bedot}
    \end{matrix}\right)
\end{align*}

We propose that the spinor variables $\lambsix$ and $\lambtsix$ in six dimensions turn into massive spinor-helicity variables in four dimensions as follows:
\begin{align}\label{six dim spinor-helicity going to massive ones}
    \lambsix^A_a \to \left( \begin{matrix}\lambda_{\alpha I} \\ -\widetilde{\lambda}^{\aldot}_I \end{matrix}\right) ~~, \qquad \qquad \lambtsix_{A\adot} \to \left( \begin{matrix}
        \lambar^\alpha_{I} \\ -{\lambtbar}_{\aldot I}
    \end{matrix} \right) ~. 
\end{align}
We can check that there is no inconsistency and the defining property of the spinor-helicity in six dimensions \eqref{defining spinor-helicity for six dim} leads us to the following relations for four-dimensional spinors:
\begin{align*}
     \psix^{AB} = \lambsix^{Aa}\, \lambsix^{Bb} \,\epsilon_{ab} = \left(\begin{matrix}
         m^* \epsilon_{\alpha\beta} & p_\alpha{}^{\bedot} \\ -p^{\aldot}{}_\beta & -m \epsilon^{\aldot \bedot}
    \end{matrix}\right) = \left( \begin{matrix}
        \lambda_\alpha^I\lambda_\beta^J\,\epsilon_{IJ} & -\lambda_\alpha^I \tilde{\lambda}^{\bedot J}\epsilon_{IJ}\\
        -\tilde{\lambda}^{\aldot I}\lambda^J_{\beta}\epsilon_{IJ} & \tilde{\lambda}^{\aldot I}\tilde{\lambda}^{\bedot J}\epsilon_{IJ}
    \end{matrix}\right)
\end{align*}
So, we have:
\begin{align}
    -p_{\alpha \bedot} = \lambda_{\alpha}^I\tilde{\lambda}_{\bedot I} \equiv |p^I]_\alpha&\langle p_I|_{\bedot} ~~, \qquad \qquad p^{\aldot \beta} = \tilde{\lambda}^{\aldot I}\lambda^{\beta}_I \equiv |p^I\rangle^{\aldot}[p_I|^\beta \label{random equation 1}\\
    m^*\epsilon_{\alpha\beta} = \lambda^I_\alpha\lambda_{\beta I} \ &\Longleftrightarrow \ |p_I]_{\beta}[p^I|^\alpha = m^*\delta_\beta^\alpha \label{square spinor mass}\\
     -m\epsilon^{\aldot \bedot} = \tilde{\lambda}^{\aldot I}\tilde{\lambda}^{\bedot}_I \ &\Longleftrightarrow \ |p^I\rangle^{\aldot}\langle p_I|_{\bedot} = m \delta^{\aldot}_{\bedot} \label{angle spinor mass}
\end{align}
Similarly, for the $\lambtsix$, we have the following relations:
\begin{align}\label{mass relation for barred}
    -p_{\alpha \bedot} &= |\barp^I]_\alpha\langle \barp_I|_{\bedot} ~~, \qquad \qquad p^{\aldot \beta} = |\barp^I\rangle^{\aldot}[\barp_I|^{\beta} \\
    |\barp_I] [\barp^I| &= m\, \mathbb{I} ~~, \qquad \qquad |\barp^I\rangle \langle\barp_I| = m^*\, \mathbb{I} 
\end{align}
The last two equations have recovered the familiar `spin-sum' relations, though a complex parameter $m$ replaces the mass. 

The `barred' variables can be solved in terms of the `unbarred' variables. Consider the quantity: $[i^I {\bari}^{K}]$, it is Lorentz invariant. However, it is not little group invariant. Since little group is only a subgroup of the Lorentz, it should be little group invariant as well, so $[i^I\bari^K] \propto \epsilon^{IK}$. We find the proportionality constant as follows:
\begin{align*}
    \begin{matrix}
        |i_I]\times  \left([i^I {\bari}{}^K] = \varphi \,\epsilon^{IK} \right) \quad \Rightarrow \quad m^*|\bari^K] = -\varphi\, |i^K] \\[6pt] 
    \left([i^I {\bari}{}^K] = \varphi \,\epsilon^{IK} \right) \times [\bari_K| \quad \Rightarrow \quad  -m[i^I| = \varphi [\bari^I|
    \end{matrix} \quad \bigg\} \quad \Rightarrow \quad \varphi = \pm \sqrt{m m^*}
\end{align*}
We can choose a particular solution $\varphi = -\sqrt{m m^*}$. Along with a similar argument for $\langle i^I\bari^K\rangle$, we have:
\begin{align}
    \label{barred variables i nterms of unbarred}
    |\bari^K] = |i^K]\sqrt{\frac{m_i}{m_i^*}} ~~~, \qquad |\bari^K\rangle = |i^K\rangle \sqrt{\frac{m_i^*}{m_i}} ~~.
\end{align}
The condition $\lambsix_i^{Aa}\lambtsix_{iA\bdot} = 0$ in six dimensions implies that $\langle i^I {\bari}^J \rangle + [i^I {\bari}^J] = 0$, and is satisfied with the above equation. Also, note that we have the following relations as well:
\begin{align*}
    \langle i^Ij^J\rangle = m \epsilon^{IJ} ~~~, \qquad [i^Ij^J] = -m^* \epsilon^{IJ}~.
\end{align*}

There exists a quadratic relation between the six-dimensional spinors, $\lambsix$ and $\lambtsix$:
\begin{align}\label{lambd in terms of lamdtilde}
    \frac{1}{2}\epsilon_{ABCD}\lambsix^{Ca}\lambsix^{Db}\epsilon_{ab} = \lambtsix_{A\adot}\lambtsix_{B\bdot}\epsilon^{\adot \bdot}.
\end{align}
In the four dimensions, it reverts back to \eqref{random equation 1}-\eqref{mass relation for barred}.

Summarizing, we express the massive four-momentum in terms of the massive spinor-helicity variables for four dimensions with the complex mass $m$ and $m^*$.  In the case of $m = m^*$, i.e., five-dimensional kinematics, we can set $|\barp^I] = |p^I]$ and $|\barp^I\rangle = |p^I\rangle$. 

\medskip
\medskip

\textbf{Lorentz Invariants:}\\
In the six-dimensional kinematics, we have
\begin{align}
    \det \lambsix_i^{Aa}\lambtsix_{jA}{}^{\bdot}= -2\psix_i.\psix_j
\end{align}
We can check that in the four dimensions, this reduces to:
\begin{align*}
    \lambsix_i^{Aa}\lambtsix_{jA}{}^{\bdot} \to -\left([i^I\overline{j}^J] + \langle i^I\overline{j}^J\rangle\right)
\end{align*}

\textbf{Three body special kinematics:} \\
The three body special kinematics in six dimensions is as follows:
\begin{align*}
    2\psix_i.\psix_j = 0 \quad \Rightarrow \quad \lambsix_i^{Aa}\lambtsix_{jA}{}^{\bdot} = u_i^a \,\tilde{u}_j^{\bdot} 
\end{align*}
The special kinematics reduces to the following statement in four dimensions:
\begin{align*}
    2p_i.p_j + m_i^*m_j + m_i m_j^* &= 0 ~~,\\
    \Rightarrow \quad \det \left([i^I\overline{j}^J] + \langle i^I\overline{j}^J\rangle\right) &= 0 \\
    \Rightarrow \qquad [i^Ij^J]\sqrt{\frac{m_j}{m_j^*}} + \langle i^Ij^J \rangle \sqrt{\frac{m_j^*}{m_j}} &\equiv u_i^I \,\tilde{u}_j^J
\end{align*}
We define the rest of the $u$ variables as follows:
\begin{align*}
    [1^I\overline{2}^J] + \langle 1^I\overline{2}^J\rangle = u_1^I \,\tilde{u}_2^J ~~&, \qquad  [2^I\overline{1}^J] + \langle 2^I\overline{1}^J\rangle = -u_2^I \,\tilde{u}_1^J ~~, \\
    [2^I\overline{3}^J] + \langle 2^I\overline{3}^J\rangle = u_2^I \,\tilde{u}_3^J ~~&, \qquad  [3^I\overline{2}^J] + \langle 3^I\overline{2}^J\rangle = -u_3^I \,\tilde{u}_2^J ~~, \\
    [3^I\overline{1}^J] + \langle 3^I\overline{1}^J\rangle = u_3^I \,\tilde{u}_1^J ~~&, \qquad  [1^I\overline{3}^J] + \langle 1^I\overline{3}^J\rangle = -u_1^I \,\tilde{u}_3^J ~~.
\end{align*}

Let us express the momentum and mass conservation in the following manner:
\begin{align*}
    \left(p_1+p_2+p_3\right)|\overline{1}_J] &+ \left(m_1+m_2+m_3\right)|\overline{1}_J\rangle = 0 \\
    \Rightarrow \quad \left(|1^I\rangle [1_I| + |2^I\rangle [2_I| + |3^I\rangle [3_I| \right)|\overline{1}_J] &+ \left(|1^I\rangle \langle1_I| + |2^I\rangle \langle 2_I| + |3^I\rangle \langle 3_I| \right)|\overline{1}_J\rangle = 0 \\
    \Rightarrow \quad -|2^I\rangle \,u_{2I}\tilde{u}_{1J} + |3^I\rangle \,u_{3I}\tilde{u}_{1J} &= 0 \\
    \Rightarrow \quad u_{2I}|2^I\rangle = u_{3I}|3^I\rangle
\end{align*}
Similar arguments tell us:
\begin{align*}
    u_{1I}|1^I\rangle = u_{2I}|2^I\rangle = u_{3I}|3^I\rangle &\equiv |u\rangle \\
    u_{1I}|1^I] = u_{2I}|2^I] = u_{3I}|3^I] &\equiv |u] \\
    \tildu_{1I}|\overline{1}^I\rangle = \tildu_{2I}|\overline{2}^I\rangle = \tildu_{3I}|\overline{3}^I\rangle &\equiv |\tildu\rangle \\ 
    \tildu_{1I}|\overline{1}^I] = \tildu_{2I}|\overline{2}^I] = \tildu_{3I}|\overline{3}^I\rangle &\equiv |\tildu\rangle  
\end{align*}
Finally, the three-body special kinematics imply the following relation:
\begin{align}\label{intersection space for six dim}
    \Aboxed{\quad  \langle \tilde{u} i^I\rangle = - [\tilde{u}i^I] ~~, \qquad \langle u\bari^I\rangle = - [u\bari^I]~. \quad } 
\end{align}
We can calculate the quantity $u_{i}^I\tildu_{iI}$ as follows:
\begin{align*}
    \left(u_1^I\tildu_{1I}\right) \left(u_{2}^J\tildu_{2J}\right) = \left(u_1^I\tildu_2^J\right)\left(-u_{2J}\tildu_{1I}\right) &= \left([1^I\overline{2}^J] + \langle 1^I\overline{2}^J\rangle \right)\left( [2_J\overline{1}_I] + \langle 2_J\overline{1}_I\rangle \right) \nonumber\\
    &= \frac{1}{\sqrt{|m_1m_2|^2}}\left(m_1m_2^*-m_1^*m_2\right)^2 
\end{align*}
Now, using the fact that $m_1m^*_2 - m_1^*m_2 = m_2m^*_3 - m_2^*m_3 = m_3m^*_1 - m_3^*m_1$, and the cyclic symmetry, we deduce that 
\begin{align}\label{u ubar contraction} 
    u_1^I\tildu_{1I} &= \frac{1}{\sqrt{m_1m_1^*}}(m_1m^*_2 - m_1^*m_2) \\
    \Rightarrow \qquad [u \tildu] &= - \langle u \tildu\rangle = m_1m^*_2 - m_1^*m_2 
\end{align}
Note that restricting to the real mass values, $u_{iI}$ and $\tildu_{iI}$ become the same. The distinction between $u_i$ and $\tildu_i$ quantifies the angles between three momenta in the compact $(p^4,p^5)$ torus.

\section{Dimensional reduction of supersymmetry algebra} \label{Appendix:C dim reduction of susy algebra}
The Dirac spinors in six dimensions have 8 components and the Weyl spinor is real: conjugation does not flip its chirality and it is its own conjugate. We are interested in the $\mathcal{N} = (1,1)$ SYM theory, having two Dirac spinors worth of supercharges. The algebra is the following: 
\begin{equation}
\begin{aligned}
    \{\qsix^{A \mathsf{I}}\,,\, \qsix^{B \sj}\} = \psix^{AB} \ \epsilon^{\si\sj} \\
    \{ \qtildesix_{A\si'}\,,\, \qtildesix_{B\sj'} \} = \psix_{AB}\ \epsilon_{\si'\sj'}
\end{aligned} 
\label{six dimensional raw algebra}
\end{equation}

$\qsix^{A(\si=1)}$ and $\qtildesix_{A(\si'=1)}$, being in the $\mathbf{4}$ and $\mathbf{4}'$ Weyl spinor representation of the six-dimensional Lorentz group $\SU(4)$ respectively, together form one Dirac spinor. The R-symmetry is $\SU(2){}^{\mathsf{I,J,\hdots}}\times \SU(2)_{\mathsf{I',J',\hdots}}$, and it mixes the two $\mathbf{4}\,$s into each other and the two $\mathbf{4}'\,$s into each other\footnote{Four-dimensional Coulomb branch theory with complex mass (equivalent to massless $\mathcal{N}=(1,1)$ theory in d=6) has R-symmetry algebra $SO(4)$. The Coulomb branch theory restricted to real mass has R-symmetry algebra $SO(5)$. Setting mass to zero in the Coulomb branch gives us back $SO(6)$ R-symmetry algebra.}.

We can express the supersymmetric algebra mentioned in \eqref{six dimensional raw algebra} using the Grassmannian coordinates $\eta_a$ and $\widetilde{\eta}_{\adot}$ as follows:
\begin{equation}
\begin{aligned}
    \qsix^{A (\si=1)} &= \lambsix^A_a\,\eta^a ~~~, \hspace{1.5cm} \qsix^{A (\si=2)} = \lambsix^A_a\,\p_{\eta_a} ~~~,\\
    \qtildesix_{A (\si'=1)} &= \lambtsix_{A\adot}\,\p_{\widetilde{\eta}_{\adot}} ~~~, \hspace{1.0cm} \qtildesix_{A (\si'=2)}=\lambtsix_{A\adot}\,{\widetilde{\eta}^{\adot}}~~.
\end{aligned} \label{eqn grassmann variables for six d}
\end{equation}
Note that $(\bullet)_{a,b,\hdots}$ and $(\bullet)_{\adot, \bdot,\hdots}$ are the six dimensional little group indices. 

Under the dimensional reduction from six dimensions to four, we have:
\begin{align*}
    \qsix^{A\si} = \left(\begin{matrix}
        Q_\alpha^\si \\
        Q^\dagger{}^{\aldot\si}
    \end{matrix}\right) ~~, \qquad \qquad \qtildesix_{A \si'} = \left(\begin{matrix}
        Q^{\alpha}_{\si'} \\
        Q^\dagger_{\aldot \si'}
    \end{matrix}\right)\ \ .
\end{align*}
Using the decomposition of $\psix$ into the four-dimensional variables as in \eqref{psix upper}, the supersymmetry algebra becomes:
\begin{align}\label{algebra 1}
    \left(\begin{matrix}
        \big\{Q_\alpha^\si\,,\, Q^\sj_\beta\big\} & \big\{Q_\alpha^\si\,,\, Q^\dagger{}^{\bedot \sj}\big\} \\ 
        \big\{ Q^\dagger{}^{\aldot \si} \,,\, Q^\sj_\beta\big\} & \big\{ Q^\dagger{}^{\aldot \si} \,,\, Q^\dagger{}^{\bedot \sj}\big\}
    \end{matrix}\right) = \left( \begin{matrix}
        m^*\epsilon_{\alpha \beta} & p_\alpha{}^{\bedot} \\
        -p^{\aldot}{}_\beta & -m\epsilon^{\aldot \bedot}
    \end{matrix} \right) \epsilon^{\si\sj} ~~~, \\
    \left( \begin{matrix}
        \big\{ Q^\alpha_{\si'} \,,\, Q^\beta_{\sj'} \big\} & \big\{ Q^\alpha_{\si'} \,,\, Q^\dagger{}_{\bedot \sj'} \big\} \\
        \big\{ Q^\dagger{}_{\aldot \si'} \,,\, Q^\beta_{\sj'} \big\} & \big\{ Q^\dagger{}_{\aldot \si'} \,,\, Q^\dagger{}_{\bedot \sj'} \big\}
    \end{matrix} \right) = \left( \begin{matrix}
        -m\epsilon^{\alpha \beta} & p^\alpha{}_{\bedot} \\
        -p_{\aldot}{}^\beta & m^*\epsilon_{\aldot \bedot} 
    \end{matrix} \right)\epsilon_{\si'\sj'} ~.
    \label{algebra 2}
\end{align}
Now, let us package the supercharges in the following manner, which is convenient for the four-dimensional perspective:
\begin{align*}
    Q_\alpha{}_{\sa} = \left(\begin{matrix}
        Q_\alpha^{\si} \\ Q_{\alpha \si'}
    \end{matrix}\right) ~~~~, \qquad Q^\dagger{}^{\aldot \sa} = \left(\begin{matrix}
        Q^\dagger{}^{\aldot}_{\si} \\
        Q^\dagger{}^{\aldot\si'}
    \end{matrix}\right)
\end{align*}
The newly introduced indices $(\bullet)^{\sa,\sbb,\hdots}$ are the $\SU(4)$ R-symmetry indices in the massless limit of the four-dimensional theory \footnote{The R-symmetry $\SU(4)$ fundamental index $(\bullet)^{\sa ,\sbb,\hdots}$ decomposes into the $\SU(2)\times \SU(2)$ spinor indices $(\bullet)^{\si,\sj,\hdots}\times (\bullet)^{\si',\sj',\hdots} $ in the same way as the 6-d Lorentz $\SU(4)$ index $(\bullet)^A$ decomposes into the 4-d spinor indices $(\bullet)^{\alpha,\beta,\hdots}\times (\bullet)^{\aldot,\bedot,\hdots} $\eqref{index decomposition}.}. Keeping the complex mass, the algebra \eqref{algebra 1} and \eqref{algebra 2} becomes: 
\begin{align*}
    \big\{ Q_{\sa\alpha} \, ,\, Q_{\sbb\beta} \big\} &= Z_{\sa\sbb}\,\epsilon_{\alpha \beta} ~~, \qquad \quad  Z_{\sa\sbb} = \left(\begin{matrix}
        m^*\Omega_{2} & 0_{2\times 2} \\
        0_{2\times 2} & -m\Omega_2
    \end{matrix}\right)_{\sa \sbb} \\
    \big\{ Q_{\sa\alpha} \, ,\, Q^\dagger{}^{\sbb\bedot} \big\} &= - p_\alpha{}^{\bedot}\,\delta_{\sa}^{\sbb} \\
    \big\{ Q^\dagger{}^{\sa\aldot} \, ,\, Q^\dagger{}^{\sbb\bedot} \big\} &= -Z^{\sa \sbb}{}\,\epsilon^{\aldot \bedot} ~~, \qquad  -Z^{\sa \sbb}{} = \left(\begin{matrix}
        -m\Omega_2 & 0_{2\times 2} \\
        0_{2\times 2} & m^* \Omega_2
    \end{matrix}\right)^{\sa\sbb} 
\end{align*}

The following is a particular realization of the supersymmetric algebra in terms of the Grassmannian variables $\eta_{iI}$ and $\etabar_{iI}$:
\begin{align*}
    Q_\alpha{}_{\sa} &= \left(\begin{matrix}
        Q_\alpha^{\si} \\ Q_{\alpha \si'}
    \end{matrix}\right) = \left(\begin{matrix}
        \left\{\begin{matrix}
            |i^I]\eta_{iI} \\ |i^I]\partial_{iI}
        \end{matrix}\right\} \\ \left\{\begin{matrix}
            |\overline{i}^I]\parbar_{iI} \\
            |\overline{i}^I]\etabar_{iI}
        \end{matrix}\right\}
    \end{matrix}\right)   ~~~~, \qquad 
    Q^\dagger{}^{\aldot \sa} = \left(\begin{matrix}
        Q^\dagger{}^{\aldot}_{\si} \\
        Q^\dagger{}^{\aldot\si'}
    \end{matrix}\right) = \left(\begin{matrix}
        \left\{ \begin{matrix}
            |i^I\rangle \p_{iI} \\ |i^I\rangle \eta_{iI}
        \end{matrix} \right\} \\ 
        \left\{ \begin{matrix}
            |\overline{i}^I\rangle \etabar_{iI} \\
            |\overline{i}^I\rangle \parbar_{iI}
        \end{matrix} \right\}
    \end{matrix}\right)
\end{align*}
Comparing it with the supersymmetric algebra in terms of Grassmannian variables in six dimensions \eqref{eqn grassmann variables for six d}, we can identify that $\eta_a \longleftrightarrow \eta_I$ and $\widetilde{\eta}_{\adot} \longleftrightarrow \widetilde{\eta}_I$. Let us emphasize that in the case of $m= m^*$, $\eta_I$ and $\etabar_I$ can be clubbed together into $\eta^a_I$, the variables used in earlier sections and in \cite{Herderschee:2019dmc,MdAbhishek:2023nvg}, retrieving a bigger $USp(4)$ R-symmetry. Note that we have:
\begin{align*}
    \{\eta_{iI}\,,\, \partial_{jK}\} = \delta_{ij}\,\epsilon_{IK} ~~~, \qquad \{\etabar_{iI}\,,\, \parbar_{jK}\} = \delta_{ij}\,\epsilon_{IK}  
\end{align*}

\section{Detailed evaluations of Grassmannian integrals from section \ref{sec:4 symplectic grassmannian integrals}}
\label{Appendix:D evaluation of the grassmannian integrals}
In this appendix, we present the detailed proof for the following equation for $n=3$ and $n=4$:
\begin{equation}
\resizebox{\textwidth}{!}{$
\begin{aligned}
    \int \frac{\dd^{n\times 2n}C}{\GL(n)}\,\delta^{n(n-1)/2}\big(C.\Omega.C^T\big)\,\delta^{2n}\,\big(C.\Omega.\Lambda^T\big)\,\delta^{2n}\big(C.\Omega.\tilde{\Lambda}^T\big) \nonumber = \delta\left(\sum \langle i^1i^2\rangle\right)\,\delta\left(\sum [ i^1i^2]\right)\,\delta^4\left(\sum |i^I\rangle [i_I|\right)
\end{aligned}$}    
\end{equation}
There are insufficient delta functions for $n>4$ to localize the integral completely and pick up residues of $f_n$.

\renewcommand{\thesubsection}{\Alph{section}.\arabic{subsection}}

\subsection{Three particle amplitude}
For $n=3$, we expect that the delta functions localize the $C$ to the following value: \eqref{C matrix}
\begin{align*}
    C_* = \left( \begin{matrix}
        \ang{q 1^1} & \ang{q 2^1} & \ang{q 3^1} & & & & \ang{q 1^2} & \ang{q 2^2} & \ang{q 3^2} \\
        \ang{u 1^1} & \ang{u 2^1} & \ang{u 3^1} & & & & \ang{u 1^2} & \ang{u 2^2} & \ang{u 3^2} \\
        [q 1^1] & [q 2^1] & [q 3^1] & & & & [q 1^2] & [q 2^2] & [q 3^2] 
    \end{matrix} \right) 
\end{align*}
Let us stress again that the upper two rows are $\tilde{\Lambda}$ and the bottom two rows are $\Lambda$ due to the fact: $\langle ui^I\rangle = - [ui^I]~.$

Since we know that \eqref{C matrix} is the solution to the delta function conditions, let us fix the following components of $C$ using the $\GL(3)$ redundancy:
\begin{align*}
C &= \left( \begin{matrix}
        \ang{q 1^1} & \ang{q 2^1} & \ang{q 3^1} & & & & c_{11} & c_{12} & c_{13}\\
        \ang{u 1^1} & \ang{u 2^1} & \ang{u 3^1} & & & & c_{21} & c_{22} & c_{23} \\
        [q 1^1] & [q 2^1] & [q 3^1] & & & & c_{31} & c_{32} & c_{33} 
    \end{matrix} \right)    \\
    &\int \frac{\dd^{2n^2}C}{\GL(n)} \hdots = (1^12^13^1)^3\int \dd^9 c_{ik} \hdots ~\bigg|_{\text{fixed }C}~.
\end{align*}
We are referring to the columns of $C$ as $1^1,2^1,3^1,{1}^2,{2}^2,{3}^2$, and we refer to the determinant of the gauge fixed part, (first $3\times 3$ minor of the $C$) as $(1^12^13^1)$. 

One can check that with the particular GL$(n)$ fixing, we have:
\begin{equation*}
\resizebox{0.9\textwidth}{!}{$
    \begin{aligned}
    \delta^{n(n-1)/2}\big(C.\Omega.C^T\big) &= \delta\left(\sum_{k=1}^3\big[c_{1k}\langle uk^1\rangle - c_{2k}\langle qk^1\rangle\big]\right)\,\delta\left(\sum_{k=1}^3\big[c_{1k}[qk^1] - c_{3k}\langle qk^1\rangle\big]\right) \nonumber \\
    &\hspace{5.2cm} \times\delta\left(\sum_{k=1}^3\big[c_{2k}[qk^1] - c_{3k}\langle uk^1\rangle\big]\right) \\
    \delta^{2n}\big(C.\Omega.\Lambda^T\big) &= \frac{1}{\langle qu\rangle^3}\,\delta^2\left(\sum_{k=1}^3 \big(|k^1\rangle c_{1k}-|k^2\rangle\langle qk^1\rangle\big)\right) \,\delta^2\left(\sum_{k=1}^3\big(|k^1\rangle c_{2k} - |k^2\rangle \langle uk^1\rangle\big)\right) \nonumber \\
    &\hspace{5cm} \times \delta^2\left(\sum_{k=1}^3\big(|k^1\rangle c_{3k} - |k^2\rangle [qk^1]\big)\right) \\
    \delta^{2n}\big(C.\Omega.\tilde{\Lambda}^T\big) &= \frac{1}{[qu]^3}\,\delta^2\left(\sum_{k=1}^3 \big(|k^1] c_{1k}-|k^2]\langle qk^1\rangle\big)\right) \,\delta^2\left(\sum_{k=1}^3\big(|k^1] c_{2k} - |k^2] \langle uk^1\rangle\big)\right) \nonumber \\
    &\hspace{5cm} \times \delta^2\left(\sum_{k=1}^3\big(|k^1] c_{3k} - |k^2] [qk^1]\big)\right)
\end{aligned}
$}
\end{equation*}
Taking care of the Jacobian factors, we have
\begin{align*}
    \int \frac{\dd^{3\times 6}C}{\GL(3)} \,\delta^{2\times 3}\big(C \ \Omega  \  \Lambda^T\big)\,\delta^{2\times 3}\big(C.\Omega.\tilde{\Lambda}^T\big) &= \delta\left([u|\sum_{k=1}^3\big(|k^1]c_{1k}-|k^2]\langle qk^1\rangle\big)\right) \nonumber \\
    &\hspace{-5cm} \times\delta\left([u|\sum_{k=1}^3\big(|k^1]c_{2k}-|k^2]\langle uk^1\rangle\big)\right)\,\delta\left([u|\sum_{k=1}^3\big(|k^1]c_{3k}-|k^2] [qk^1]\big)\right) \,\bigg|_{c_{ik}=c_{*ik}}~.
\end{align*}
The solutions ($c_{*ik}$) on the support of delta functions, as expected, are as follows:
\begin{align*}
    c_{*1k}{} = \langle qk^2\rangle ~, \qquad c_{*2k} = \langle uk^2\rangle ~, \qquad c_{*3k} = [qk^2] ~.
\end{align*}
Evaluating the remaining delta functions at these solutions, we obtain that $C$ is determined as \eqref{C matrix}, and we have:
\begin{equation}
\resizebox{0.91\textwidth}{!}{$
\begin{aligned}
    &\int \frac{\dd^{3\times 6}C}{\GL(3)}\ \delta^{3}\big(C . \Omega .   C^T\big) \,\delta^{2\times 3}\big(C.\Omega.\Lambda^T\big)\,\delta^{2\times 3}\big(C.\Omega.\tilde{\Lambda}^T\big) = \delta\left(\sum \langle i^1i^2\rangle\right)\delta\left(\sum [i^1i^2]\right)\,\delta^4\left(P\right). \label{n=3 integral carried out}
\end{aligned}$}
\end{equation}
Recall that $\langle qu\rangle=1$ and $[qu]=1$~.

\subsection{Four particle amplitude}
We have the following expected solution to the delta functions constraining the $C$ for $n=4$:
\begin{align*}
    \text{Expected } \quad C_* = \left( \begin{matrix}
        \Lambda_{2\times 2n} \\ \tilde{\Lambda}_{2\times 2n}
    \end{matrix}\right)_{4\times 2n}
\end{align*}
We shall see that this is indeed the case. Recall that $\Lambda$, and $\tilde{\Lambda}$ are of the following form:
\begin{align*}
    \Lambda &= \left(
    \begin{matrix} \langle 1^1| & \hdots \langle n^1| & | & \langle 1^2 | & \hdots \langle n^2| \end{matrix}\right)_{2\times 2n}   \\
    \tilde{\Lambda} &= \left(
    \begin{matrix} |1^1] & \hdots |n^1] & | & |1^2]  & \hdots |n^2] \end{matrix}\right)_{2\times 2n}
\end{align*}
Motivated by this, let us fix the $\GL(n)$ redundancy of $C$ as follows:
\begin{align*}
    &C = \left(\begin{matrix}
        \left(\begin{matrix}|1_2]_b & \hdots & |4_2]_b \end{matrix}\right) & \left(\begin{matrix}{\tc}_{b1} & \hdots & {\tc}_{b4} \end{matrix}\right) \\
        \left(\begin{matrix}\langle 1_2|_{\bd} & \hdots & \langle 4_2|_{\bd} \end{matrix}\right) & \left(\begin{matrix}c_{\bd 1 } & \hdots & c_{\bd 4} \end{matrix}\right)
    \end{matrix}\right)_{4\times 8} = \left(\begin{matrix}
        \{|i_2]_b\}_{2\times 4}& \qquad  & \{\tc_{bi}\}_{2\times 4} \\
        \{\langle i_2|_{\bd}\}_{2\times 4}& \qquad  & \{c_{\bd i}\}_{2\times 4}
    \end{matrix}\right) \\
    &\hspace{2cm}\int \frac{\dd^{4\times 8}C}{\GL(4)} = (1^12^13^14^1)^4\int \dd^{4\times 4}c_{ik}~.    
\end{align*}
Note that the subscripts $b$ and $\bd$ are the spinor indices. We shall suppress them in certain variables to avoid clutter. The maximal minor $(1^12^13^14^1)$ is the determinant of the gauge-fixed part of $C$. 

The matrix $C.\Omega.C^T$ is antisymmetric, and there are six independent constraints:
\begin{equation}\label{C.O.CT}
    \resizebox{0.9\textwidth}{!}{$
    \begin{aligned}
    \delta^{n(n-1)/2}\big(C.\Omega.C^T\big) = \delta\left( \sum_i [i_2|^a\,\tilde{c}_{ai} \right)  \delta\left( \sum_ic_{\dot{a}i}\,|i_2\rangle^{\dot{a}} \right)  \delta^4\left(  \sum_i -|i_2]_b\,c_{\ad i} + {\tc}_{bi}\langle i_2|_{\ad}\right)
    \end{aligned}
    $}
\end{equation}
We compute $C.\Omega.{\Lambda}$ and $C.\Omega.\tilde{\Lambda}$, and obtain:
\begin{align*}
    \delta^8\big(C.\Omega.\Lambda^T\big)\ \delta^8\big(C.\Omega.\tilde{\Lambda}^T\big) = &\delta^4\left(\sum_i|i_2]_b \langle i_1| + \tc_{bi}\langle i_2|\right)\delta^4\left(\sum_i \langle i_2|_{\bd} \langle i_1| + c_{\bd i}\langle i_2|\right) \nonumber \\
    &\quad \times \delta^4\left(\sum_i|i_2]_b |i_1] + \tc_{bi} |i_2]\right)\delta^4\left(\sum_i \langle i_2|_{\bd} |i_1] + c_{\bd i} |i_2]\right) \nonumber\\
    &= \delta^4\left(\sum_i|i_2]_b \langle i_1| + \tc_{bi}\langle i_2| - \langle i_2|_{\bd} |i_1] - c_{\bd i} |i_2]\right) \nonumber \\
    &\hspace{-4.5cm} \times \delta^4\left(\sum_i \langle i_2|_{\bd} \langle i_1| + c_{\bd i}\langle i_2|\right) \delta^4\left(\sum_i|i_2]_b |i_1] + \tc_{bi} |i_2]\right)\delta^4\left(\sum_i \langle i_2|_{\bd} |i_1] + c_{\bd i} |i_2]\right)
\end{align*}
On the support of the $\delta^4$ from the constraint $C.\Omega.C^T = 0$, \eqref{C.O.CT}, the first delta function becomes:
\begin{equation*}
\resizebox{\textwidth}{!}{$
\begin{aligned}
    &\delta^8\big(C.\Omega.\Lambda^T\big)\ \delta^8\big(C.\Omega.\tilde{\Lambda}^T\big) = \delta^4\left(\sum_i|i^I] \langle i_I| \right)  \\
    &\qquad \times \delta^4\left(\sum_i \langle i_2|_{\bd} \langle i_1| + c_{\bd i}\langle i_2|\right) \delta^4\left(\sum_i|i_2]_b |i_1] + \tc_{bi} |i_2]\right)\delta^4\left(\sum_i \langle i_2|_{\bd} |i_1] + c_{\bd i} |i_2]\right)
\end{aligned}$}
\end{equation*}
Combining everything together, we have:
\begin{equation*}
\resizebox{\textwidth}{!}{$
\begin{aligned}
    &\int \frac{\dd^{n\times 2n}C}{\GL(n)}\,\delta^{n(n-1)/2}\big(C.\Omega.C^T\big)\,\delta^{2n}\,\big(C .\Omega .  \Lambda^T\big)\,\delta^{2n}\big(C .\Omega  .\tilde{\Lambda}^T\big)  \\
    &= \delta^4\big(P\big)\, (1^12^13^14^1)^4\int \dd^{8}c_{\ad i}\,\dd^8\tc_{ai}\ \delta\left( \sum_i [i_2|^a\,\tilde{c}_{ai} \right)  \delta\left( \sum_ic_{\dot{a}i}\,|i_2\rangle^{\dot{a}} \right)  \delta^4\left(  \sum_i -|i_2]_b\,c_{\ad i} + {\tc}_{bi}\langle i_2|_{\ad}\right)  \\
    &\qquad \times \delta^4\left(\sum_i \langle i_2|_{\bd} \langle i_1| + c_{\bd i}\langle i_2|\right) \delta^4\left(\sum_i|i_2]_b |i_1] + \tc_{bi} |i_2]\right)\delta^4\left(\sum_i \langle i_2|_{\bd} |i_1] + c_{\bd i} |i_2]\right) 
\end{aligned}$}
\end{equation*}
The integrals over $c$ and $\tilde{c}$ are carried out as follows:
\begin{align*}
    \int \dd^8 c_{\ad i}\ \delta^4\left(\sum_i \langle i_2|_{\bd} \langle i_1| + c_{\bd i}\langle i_2|\right) \delta^4\left(\sum_i \langle i_2|_{\bd} |i_1] + c_{\bd i} |i_2]\right) &= \left(\frac{1}{4}\sum_{i,j,k,l}\ang{i_2j_2}[k_2l_2]\,\epsilon^{ijkl}\right)^{-2} \nonumber \\ 
    &= (1^12^13^14^1)^{-2}~.
\end{align*}
Note that the factor on the RHS is the minor of the $C$ matrix corresponding to the gauge-fixed part. A total of $(1^12^13^14^1)^{-4}$ is produced as a result of the integral, which cancels appropriately with the factors present already. The $c$ and $\tilde{c}$ are completely determined as $c_{\ad i} = \langle i_1|_{\bd}$ and $\tilde{c}_{bi} = |i_1]_b$. So, we obtain the expected solution for $C$. 

After the integration, the delta functions $\delta\left( \sum [i_2|^a\,\tilde{c}_{ai} \right)  \delta\left( \sum c_{\dot{a}i}\,|i_2\rangle^{\dot{a}} \right) $ simply become $\delta\big(\sum \langle i^1i^2\rangle\big)$ $\delta\big(\sum  [i^1i^2]\big)$. So, for $n = 4$, we have established that:
\begin{equation*}
\resizebox{\textwidth}{!}{$
\begin{aligned}
    &\int \frac{\dd^{n\times 2n}C}{\GL(n)}\,\delta^{n(n-1)/2}\big(C.\Omega.C^T\big)\,\delta^{2n}\,\big(C.\Omega.\Lambda^T\big)\,\delta^{2n}\big(C.\Omega.\tilde{\Lambda}^T\big) = \delta^4(P) \,\delta\left(\sum \ang{i_1i_2}\right)\delta\left(\sum[i_1i_2]\right)  \\
    &\hspace{4cm} 
    \text{and} \quad C_* = \left(\begin{matrix}
        \Lambda \\ \tilde{\Lambda}
    \end{matrix}\right)
\end{aligned}$}
\end{equation*}
\section{Equivalence of three-point amplitude of six-dimensional theory} \label{Appendix:E equivalent of three-point amplitude}
In this section, we show how the three-point amplitude for six-dimensional maximal supersymmetric YM \eqref{three-point amplitude for six dimensional theory}  matches the form in \cite{Dennen:2009vk}. We have expressed the amplitude as follows:
\begin{align*}
    &\mathcal{A}_3 = \delta^3\left(C_*  .
 \Omega  . 
 \eta^T\right)\,\delta^3\left(\overline{C}_* .   \Omega  . \etabar^T\right)~~, \\
    C_*&\equiv  \left(\begin{matrix}
        \langle \tilde{q} i^I\rangle \\ \langle \tilde{u} i^I\rangle \\ [\tilde{q} i^I]
    \end{matrix}\right)_{3\times 6} ~~, \qquad \overline{C}_* = \left( \begin{matrix}
        \langle q\bari^I\rangle  \\ \langle u\bari^I\rangle \\ [q\bari^I]
    \end{matrix} \right)_{3\times 6} ~.
\end{align*}
The $\tilde{u}$ and ${u}$ variables encode the intersection space of $\lambsix$ and $\lambtsix$ respectively, and $|q\rangle~,\hdots$ are defined as $\langle qu\rangle=1~, \hdots~.$ 

The variables $w_i^I$ is defined as follows: $u_{iI}w_i^I = 1$ \cite{Dennen:2009vk}. Analogous to the $|u\rangle$, we can define spinor $|i^w\rangle = w_{iI}|i^I\rangle$ and $|i^w] = w_{iI}|i^I]~.$ The ambiguity in the definition of $w_i^I$ is partially fixed by demanding that $\sum|i^w\rangle = 0 = \sum |i^w]~.$ Let us write the supersymmetric delta functions using this special frame defined by $u_{iI}$ and $w_{iI}$:
\begin{align*}
    |i^I\rangle\eta_{iI} = |i^I\rangle\,u_{i,J}w^{J}_i\,\eta_{iI} = -\left(|i_J\rangle\,u_i^Jw_i^I + |i^J\rangle\,u_i^Iw_{iJ}\right)\eta_{iI} \equiv |u\rangle \,\eta_{iw} - |i^w\rangle \,\eta_{iu}~.
\end{align*}
We have introduced a shorthand $\eta_{iw} = w_i^{I}\eta_{iI}$ and $\eta_{iu} = u_{i}^I\eta_{iI}$ analogous to \cite{Herderschee:2019dmc}. For the square variables, we have:
\begin{align*}
    |i^I]\,\eta_{iI} = |u] \,\eta_{iw} - |i^w]\,\eta_{iu}~.
\end{align*}
 The two relations $[\tilde{u}u] = - \langle \tilde{u}u\rangle$ (see \eqref{u ubar contraction}) and $[\tilde{u}i^w] = -\langle \tilde{u}i^w\rangle$, imply  $\langle \tilde{u}i^I\rangle\eta_{iI} = - [\tilde{u}i^I]\eta_{iI}$. 

Let us focus on the $\delta^3\left(C_* . \Omega.\eta^T\right)$ for now:
\begin{align*}
    \delta^3\left(C_* . \Omega . \eta^T\right) = \delta\left(\sum \langle \tilde{q}i^I\rangle\eta_{iI}\right)\,\delta\left(\sum \langle \tilde{u}i^I\rangle\eta_{iI}\right)\,\delta\left(\sum [ \tilde{q}i^I]\eta_{iI}\right)
\end{align*}
In the special little group frame defined by $u$ and $w$ variables, we have,
\begin{align*}
    \delta^3\left(C_* . \Omega . \eta^T\right) &= \delta \left( \langle \tilde{q}u\rangle\sum \eta_{iw} - \sum \langle \tilde{q}i^w\rangle\eta_{iu} \right) \ \delta \left( \langle \tilde{u}u\rangle\sum \eta_{iw} - \sum \langle \tilde{u}i^w\rangle\eta_{iu} \right) \nonumber\\
    &\hspace{3cm} \times \delta\left( [\tilde{q}u]\sum\eta_{iw} - \sum[\tilde{q}i^w]\eta_{iu} \right)
\end{align*}
Let us substitute the solution for $\sum \eta_{iw}$ from second delta function into the other two:
\begin{equation*}
    \resizebox{\textwidth}{!}{$
    \begin{aligned}
    \delta^3\left(C_* . \Omega . \eta^T\right) &= \delta \left( \langle \tilde{q}u\rangle \left[\frac{1}{\langle\tilde{u}u\rangle}\sum \langle \tilde{u}i^w\rangle\eta_{iu}\right] - \sum \langle \tilde{q}i^w\rangle\eta_{iu} \right) \ \delta \left( \langle \tilde{u}u\rangle\sum \eta_{iw} - \sum \langle \tilde{u}i^w\rangle\eta_{iu} \right) \nonumber\\
    &\hspace{3cm}\times \delta\left( [\tilde{q}u]\left[\frac{1}{[\tilde{u}u]}\sum [\tilde{u}i^w]\eta_{iw}\right] - \sum [\tilde{q}i^w]\eta_{iu} \right) \\
    &\hspace{-1cm}= \frac{1}{\langle \tilde{u}u\rangle[\tilde{u}u]} \delta\left(\sum \langle ui^w\rangle\eta_{iu}\right)\ \delta \left( \langle \tilde{u}u\rangle\sum \eta_{iw} - \sum \langle \tilde{u}i^w\rangle\eta_{iu} \right)\ \delta\left(\sum [ ui^w]\eta_{iu}\right)
\end{aligned}$}
\end{equation*}
Let us use the fact that for anti-commuting quantities $\delta(\eta) = \eta$.
\begin{equation*}
    \resizebox{\textwidth}{!}{$
    \begin{aligned}
    \delta^3\left(C_* . \Omega.\eta^T\right) &= \frac{1}{\langle \tilde{u}u\rangle[\tilde{u}u]} \left(\sum \langle ui^w\rangle\eta_{iu}\right)\left(\sum [ ui^w]\eta_{iu}\right)\left( \langle \tilde{u}u\rangle\sum \eta_{iw} - \sum \langle \tilde{u}i^w\rangle\eta_{iu} \right) \nonumber\\
    &= \frac{1}{\langle \tilde{u}u\rangle[\tilde{u}u]}\bigg[\left( \langle u1^w\rangle[u2^w] - \langle u2^w\rangle[u1^w] \right)\,\eta_{1u}\eta_{2u} + \text{cyclic} \bigg] \times \left( \langle \tilde{u}u\rangle\sum \eta_{iw} - \sum \langle \tilde{u}i^w\rangle\eta_{iu} \right).
    \end{aligned}$}
\end{equation*}
Now, using the fact that $[ui^w] = -m_i^*$ and $\langle ui^w\rangle = m_i$, we deduce that 
\begin{align*}
\langle u1^w\rangle[u2^w] - \langle u2^w\rangle[u1^w] = -m_1m_2^* + m_1^*m_2 = [ \tilde{u}u]~.
\end{align*}
Similar identities hold for the cyclic permutations of LHS. Thus, we are led to the following:
\begin{align*}
    \Rightarrow \quad \delta^3\left(C_* . \Omega.\eta^T\right) &= \frac{1}{\langle \tilde{u}u\rangle}\big(\eta_{1u}\eta_{2u}+\eta_{2u}\eta_{3u}+\eta_{3u}\eta_{1u} \big)\left( \langle \tilde{u}u\rangle\sum \eta_{iw} - \sum \langle \tilde{u}i^w\rangle\eta_{iu} \right)
\end{align*}
The term with $\langle \tilde{u}i^w\rangle\eta_{iu}$ vanish due to the fact that $\sum |i^w\rangle = 0$, and we are left with the following:
\begin{align*}
    \delta^3\left(C_* . \Omega.\eta^T\right) = \big(\eta_{1u}\eta_{2u}+\eta_{2u}\eta_{3u}+\eta_{3u}\eta_{1u} \big)\left(\sum\eta_{iw}\right)~.
\end{align*}
A similar analysis for the $\overline{C}_*$ sector gives us the final amplitude:
\begin{align*}
    \mathcal{A}_3 &= \delta^3\left(C_* . \Omega.\eta^T\right)\,\delta^3\left(\overline{C}_* . \Omega.\etabar^T\right)\nonumber \\
    &= \big(\eta_{1u}\eta_{2u}+\eta_{2u}\eta_{3u}+\eta_{3u}\eta_{1u} \big)\left(\sum\eta_{iw}\right) \times \left(\etabar_{1\tilde{u}}\etabar_{2\tilde{u}} + \etabar_{2\tilde{u}}\etabar_{3\tilde{u}} + \etabar_{3\tilde{u}}\etabar_{1\tilde{u}}\right)\left(\sum \etabar_{i\tilde{w}}\right)~.
\end{align*}
Note that we have $\etabar_{i\tilde{u}} = \tilde{u}^{iI}\etabar_{iI}$ and $\etabar_{i\tilde{w}} = \tilde{w}^{iI}\etabar_{iI}~.$ The final form matches with the one in \cite{Dennen:2009vk}. One can find a particular component amplitude, three gluon amplitude and obtain the following familiar function due to \cite{Cheung:2009dc}:
\begin{align*}
    A_3\left( 1_{a\adot}\, 2_{b\bdot} \, 3_{c\dot{c}}\right) &= \Gamma_{abc}\Gamma_{\adot \bdot \dot{c}} \nonumber\\
    &\hspace{-1cm}= \left(u_{1a}u_{2b}w_{3c} + u_{1a}w_{2b}u_{3c} + w_{1a}u_{2b}u_{3c}\right)\left(\tilde{u}_{1\adot}\tilde{u}_{2\bdot}\tilde{w}_{3\dot{c}} + \tilde{u}_{1\adot}\tilde{w}_{2\bdot}\tilde{u}_{3\dot{c}} + \tilde{w}_{1\adot}\tilde{u}_{2\bdot}\tilde{u}_{3\dot{c}}\right)~.
\end{align*}
In the last line, we have lifted the four-dimensional (massive) little group to the little group of six-dimensional theory. 

\bibliographystyle{JHEP}
\bibliography{amplitudes}

\end{document}